\documentclass[aps,prl,twocolumn,floats,epsfig,pdflatex]{revtex4-2}
\usepackage{graphicx}
\usepackage{bbm}
\usepackage{physics}
\usepackage{amsthm}
\usepackage{amssymb} 
\usepackage{amsmath}
\usepackage{color}
\usepackage{titlesec}
\usepackage[normalem]{ulem}
\usepackage{enumitem}
\usepackage{pifont}
\usepackage{cancel}
\usepackage[caption=false]{subfig}
\usepackage[colorlinks=true, citecolor=blue, urlcolor=blue, linkcolor=magenta, pdfencoding=auto, psdextra]{hyperref}
\usepackage[all]{hypcap}
\usepackage{mathtools}
\usepackage{tikz}
\usepackage{adjustbox}
\usepackage{makecell}
\usepackage{hyperref}
\usepackage[normalem]{ulem}

\titleformat{\section}[runin]
            {\normalfont\itshape}{\thesection}{1em}{}[---]
\titlespacing*{\section}
              {1em}
              {0.3em}
              {0.0em}

\titleformat{\subfigure}[leftmargin]{}{}{0pt}{}
              
\setlist{nosep}
\newcommand{\dummylabel}[2]{%
  \begingroup
    \edef\@currentlabel{#2}%
    \protected@write\@auxout{}{%
      \string\newlabel{#1}{{#2}{\thepage}{}{}{}}%
    }%
  \endgroup
}

\makeatletter

\def\maketitle{
\@author@finish
\title@column\titleblock@produce
\suppressfloats[t]}
\makeatother

\begin{document}
\title{Dressed Floquet scars from protected zero-quasienergy modes in a Rydberg chain} 
\author{Saptadip Roy\textsuperscript{1},
Bhaskar Mukherjee\textsuperscript{2,3},
K.~Sengupta\textsuperscript{1} and
Arnab Sen\textsuperscript{1}} 
\affiliation{{\bf 1} School of Physical Sciences, Indian Association for the Cultivation of Science, Kolkata 700032, India \\
{\bf 2} School of Physics, University of Hyderabad, Prof. C. R. Rao Road, Gachibowli, Hyderabad 500046, India \\
{\bf 3} S. N. Bose National Centre for Basic Sciences, Block JD, Sector III, Salt Lake, Kolkata 700106, India}

\date{\today}
\begin{abstract}

We show that a symmetry-protected Floquet nullspace can host anomalous many-body eigenstates that are obtained by projecting structured parent states onto an exact zero-quasienergy sector. We demonstrate this in a periodically driven PXP chain with an exponentially large (in system size) protected nullspace. Over the accessible system sizes, special parent states show an exponentially increasing enhancement of their nullspace weight relative to the typical value, even as their absolute overlaps decrease. The resulting Floquet scars can therefore become globally orthogonal to their parents while retaining anomalous correlations inherited from them. We demonstrate this mechanism for two contrasting parents: a locally rotated volume-entangled Ivanov–Motrunich scar and the unentangled Rydberg vacuum. Floquet perturbation theory captures the weak-dressing regime, while exact diagonalization shows persistence into nonperturbative regimes where the Floquet Hamiltonian generically contains increasingly nonlocal terms. Our results identify projection into a protected zero-quasienergy sector as a route to Floquet quantum many-body scars.

\end{abstract}

\maketitle
\phantomsection

\section*{Introduction}

The eigenstate thermalization hypothesis (ETH) posits that individual finite-energy-density eigenstates of a nonintegrable Hamiltonian reproduce thermal expectation values of few-body observables~\cite{Deutsch1991, Srednicki1994, Rigol2008, dAlessio2016}. Quantum many-body scars (QMBSs) are striking exceptions: a sparse set of atypical eigenstates remains strongly athermal while embedded in an otherwise ETH-respecting spectrum~\cite{Moudgalya2018, Moudgalya_2_2018, LinMotrunich2019, Bull2019, Schecter2019, Moudgalya2020a, Moudgalya2020b, McClarty2020, Mohapatra2023, Desaules2023a, Desaules2023b, Giudici2024, Kerschbaumer2025}. Initial states with anomalously large overlap on such eigenstates can retain memory and exhibit weak ergodicity breaking, unlike many-body localization where thermalization fails for typical states~\cite{MBLreview}. The observation of long-lived revivals from a N\'eel state in a $51$-atom Rydberg chain~\cite{Bernien2017}, subsequently understood through constrained dynamics~\cite{Turner2018pxp1, Turner2018pxp2} of the PXP model~\cite{Sachdev_pxp, Fendley_pxp, Lesanovsky_pxp}, has made the origin and stability of such exceptional states a central problem in nonequilibrium quantum matter.

Several distinct mechanisms for generating QMBSs are now known~\cite{Shiraishi2017, Choi2019, Khemani2019, Iadecola2019, Sugiura2021, Serbyn2021PXPscars, Mukherjee_fragmentation, Moudgalya_2022_review, Chandran2023review, Moudgalya2023, Omiya2023a, Omiya2023b, Moudgalya2024, sharma2026, tan2025interferencecagedquantummanybodyscars, Jonay2026, benami2025manybodycagesdisorderfreeglassiness, Sashi_Fockspacecages}, but a general principle identifying which states remain exceptional is still lacking. A particularly intriguing setting arises when chiral and point-group symmetries intertwine~\cite{Banerjee2021, Karle2021}, enforcing through an index theorem an exponentially large number (in system size) of exact mid-spectrum zero modes in an interacting system with an $E\leftrightarrow -E$ symmetric spectrum~\cite{Turner2018pxp1, Turner2018pxp2, Schecter2018_index}. A generic state in this degenerate manifold is expected to be locally thermal according to ETH~\cite{Schecter2018_index}. Nevertheless, structured zero-energy scars have been identified inside protected nullspaces of local Hamiltonians, from constrained spin chains to lattice gauge theories~\cite{Biswas2022, Udupa2023, Sau2024, Budde2024, Pal2025, Ivanov2025, ivanov2025exactarealawscareigenstates, Gupta2026, Shashikanta2025, Eloi2026, Mukherjee2026}. 

Whether an analogous mechanism survives in Floquet matter is nontrivial. In a periodically driven system with a time-varying Hamiltonian $H(t)=H(t+T)$, where $T$ denotes the time-period of the drive, the fundamental spectral object is the Floquet unitary $U(T,0)=\mathcal T\exp\!\left[-i\int_0^T H(t)\,dt\right]=\exp(-iH_F T)$ where $\mathcal T$ denotes time ordering and $H_F$ represents the Floquet Hamiltonian~\cite{Bukovreview2015, Eckardt2017, Moessnerreview2017}. Although $H(t)$ is local, the Floquet Hamiltonian, $H_F$, generally contains an infinite series of increasingly nonlocal operators~\cite{Lazarides2014, Alessio2014, Ponte2015, Abanin2015, Kuwahara2016, Mori2016}. Moreover, the quasienergies (eigenvalues of $H_F$) are defined modulo $2\pi/T$, and both $0$ and $\pi/T$ are fixed points of quasienergy reflection. Thus, a static protected nullspace construction does not carry over automatically.
For generic interacting Floquet systems, Floquet ETH predicts that typical quasienergy eigenstates reproduce infinite-temperature expectation values of few-body observables~\cite{Alessio2014, Lazarides2014}.
This raises two related questions. First, can a symmetry-protected Floquet nullspace host anomalous zero-quasienergy states analogous to scars in nullspaces of local Hamiltonians? Second, how can one distinguish such a scarred state from a typical state inside an exponentially large protected nullspace?

In this Letter, we answer these questions in a periodically driven PXP chain~\cite{Mukherjee2020a, Mukherjee2020b, Hudomal2022} whose Floquet Hamiltonian possesses an exponentially large (in system size) symmetry-protected zero-quasienergy subspace. For a parent state $|\psi\rangle$, we define $W_0(\psi)=\langle \psi|\mathbb{P}_0|\psi \rangle$, where $\mathbb{P}_0$ projects onto the zero-quasienergy subspace, and construct a zero-quasienergy mode $|Z_\psi \rangle = \mathbb{P}_0 |\psi \rangle/\sqrt{W_0(\psi)}$. The nontrivial question is whether suitable parents exist with anomalously large projection weight compared to typical states in the same symmetry sector over a range of drive parameters. We therefore introduce the typical nullspace weight $W_0^{\mathrm{typ}}$ and use \(W_0/W_0^{\rm typ}\gg1\) as a central diagnostic of an anomalously selected zero-quasienergy state, establishing its scarred character through the persistence of parent-specific few-body correlations, including system-spanning ones. Our results are consistent with such a Floquet scar becoming globally orthogonal to its parent as the system size increases, while retaining an exponentially enhanced projection weight relative to a typical zero-quasienergy mode.

\begin{figure}[t] \label{fig_1_PRL_resub}
\includegraphics[width=0.9\linewidth]{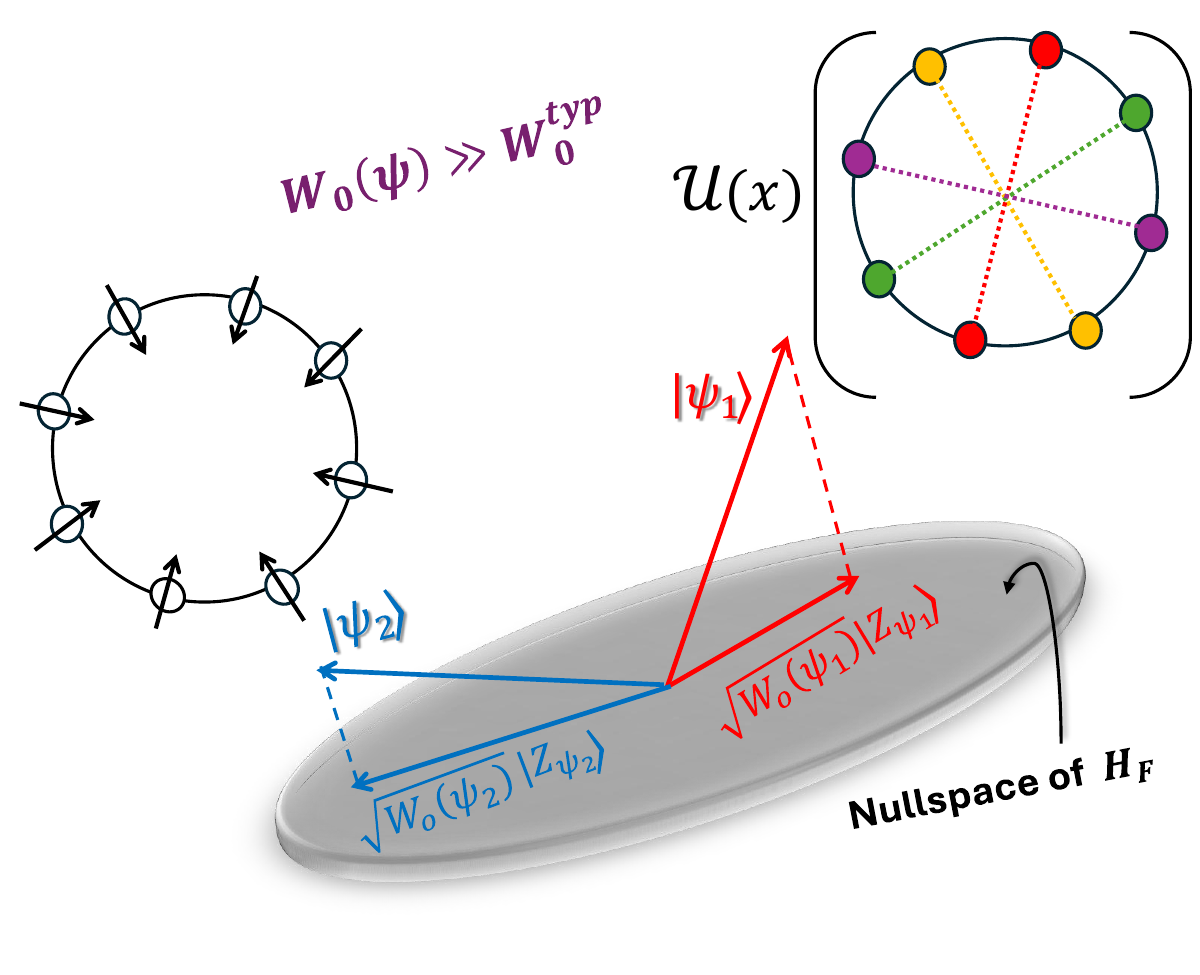}
\caption{Schematic representation of the anomalous zero-quasienergy modes, $|Z_{\psi} \rangle$ (Eq.~\ref{eq:defdescendant}), in the nullspace of the Floquet Hamiltonian $H_F$ (Eq.~\ref{eq3}). These modes can be viewed as projections of two parent states onto this nullspace, with both having anomalously large projection weight compared to typical states ($W_0^{\mathrm{typ}}$). \(|\psi_1\rangle\) represents the highly entangled Ivanov–Motrunich scar after the local unitary rotation \(\mathcal{U}(x)\) of Eq.~\ref{rotatedIM}; its antipodal spins remain perfectly correlated. $|\psi_2 \rangle$ represents the unentangled Rydberg vacuum state. }
\end{figure}

We establish this mechanism for two radically different parent states. The first is a local unitary rotation of the volume-law-entangled Ivanov--Motrunich (IM) zero-mode scar of the undriven PXP model~\cite{Ivanov2025}, whose antipodal spins are perfectly correlated (Fig.~\ref{fig_1_PRL_resub}). The second is the completely unentangled Rydberg vacuum, which rapidly thermalizes under the undriven PXP Hamiltonian~\cite{Turner2018pxp1, Turner2018pxp2}. Previous work~\cite{Mukherjee2020b} found near-perfect dynamical freezing of this state at isolated drive frequencies in the high-drive-amplitude regime. Here the role of the vacuum is qualitatively different: projection onto the symmetry-protected Floquet nullspace produces an anomalous exact zero-quasienergy eigenstate over an extended range of parameters, including regimes in which the freezing mechanism and low-order Floquet perturbation theory have broken down.

\begin{figure*} \label{fig2_PRL_resub}
    \includegraphics[width=0.48\linewidth]{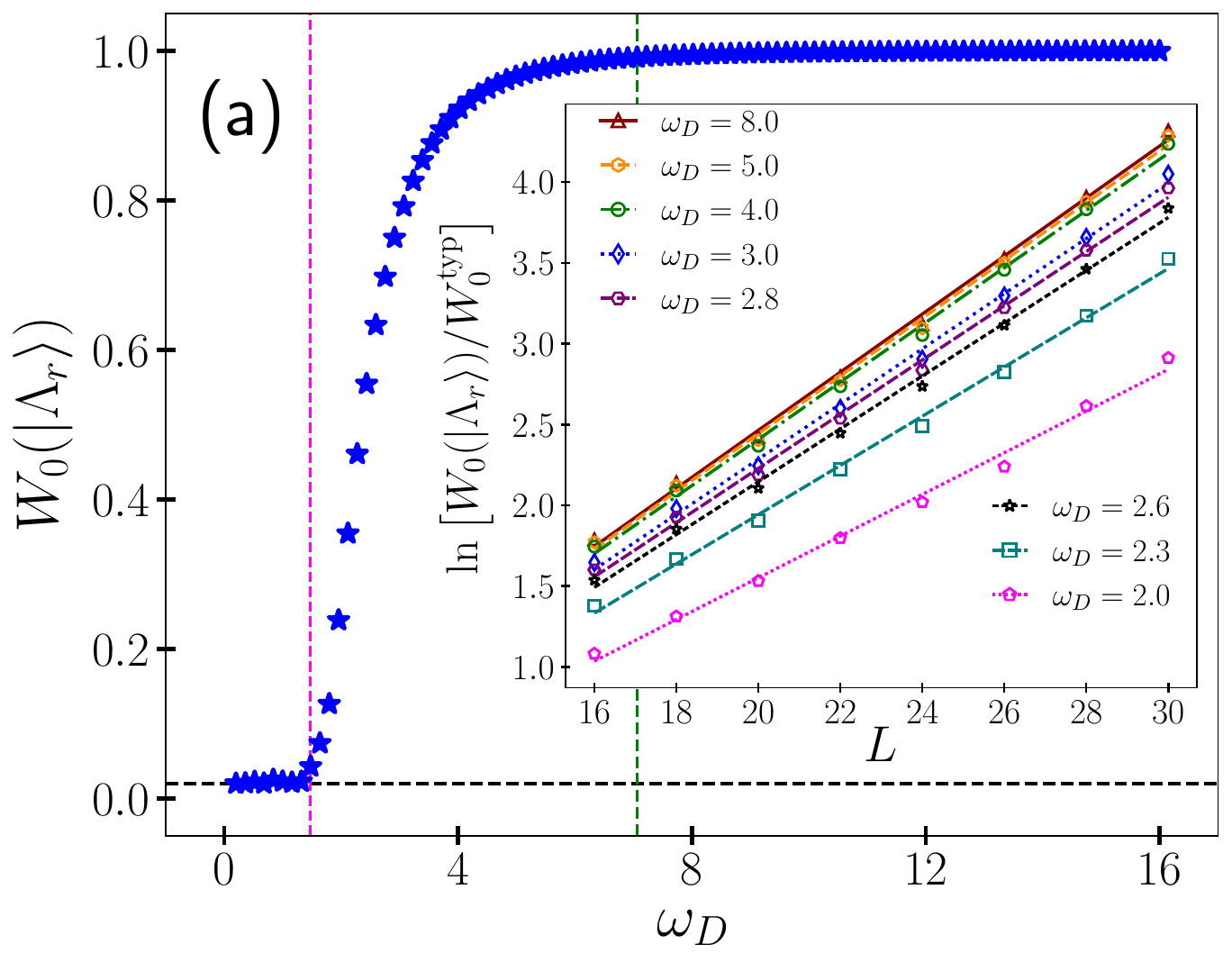} 
    \includegraphics[width=0.48\linewidth]{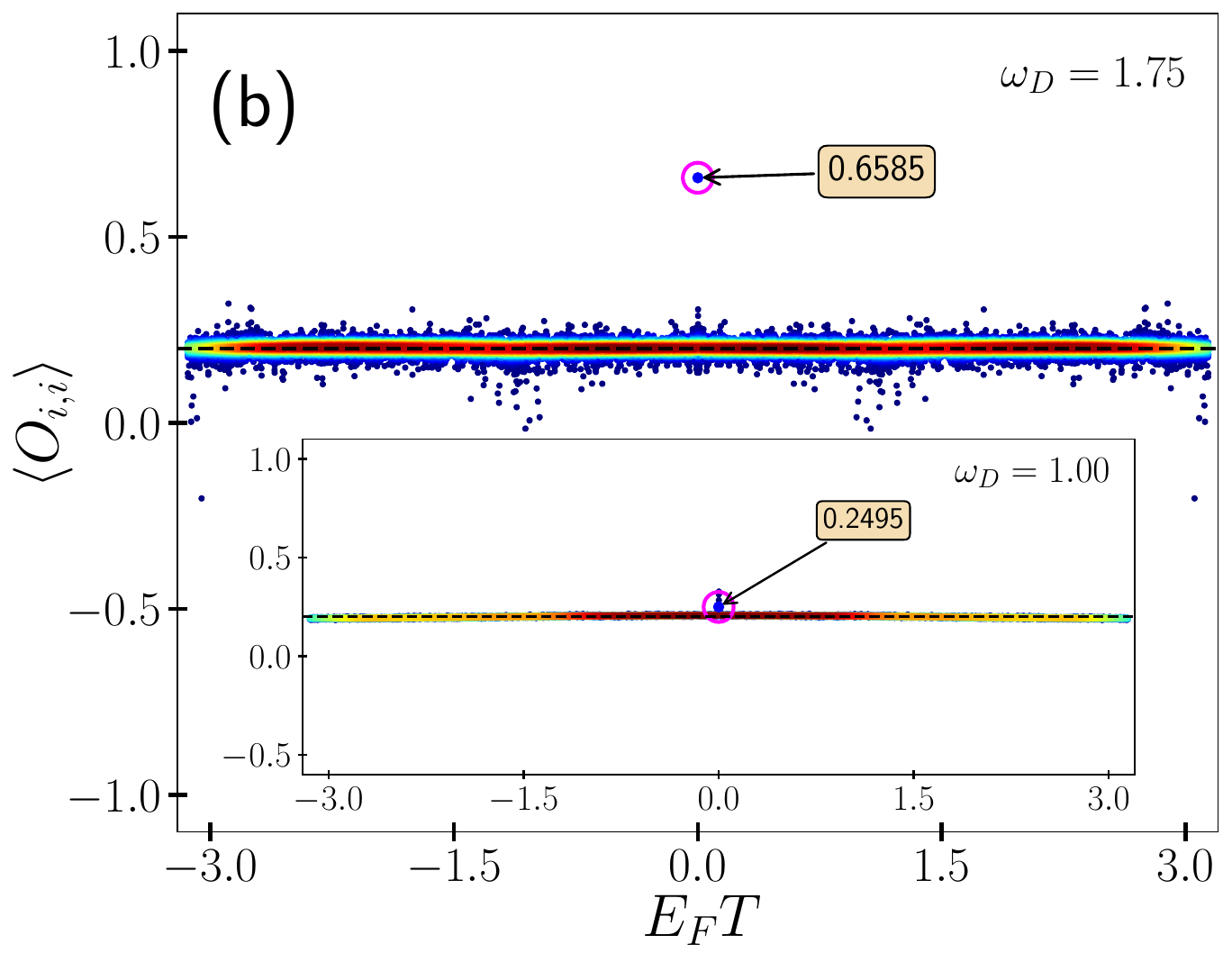} 
    \caption{(a) $W_0(\ket{\Lambda_r})$ for $L=28$ at $x=1.2$ as a function of $\omega_D$. The vertical dashed lines (magenta and green) mark the onset of the two plateaus of $W_0(\ket{\Lambda_r})$, while a horizontal black dashed line shows $W_0^{\rm typ} \approx 0.02$ for $L=28$. The inset shows $\ln \left[W_0(\ket{\Lambda_r})/ W_0^{\rm typ}\right]$ as a function of $L$ for $x=1.2$ and several $\omega_D$ near the knee regime. (b) $\langle \mathcal{O}_{i,i}\rangle$ for the Floquet eigenstates ${\ket{E_F}}$ for $L=30$ in the $K=0,\mathcal{P}=+1$ sector. The density of states is indicated by the same color map in both panels, where warmer colors correspond to higher density of states. The state $\ket{Z_{\Lambda_r}}$ is highlighted in magenta, and the corresponding correlator value is indicated in both panels of (b). $W_0(\ket{\Lambda_r})/W_0^{\rm typ} \approx 6.93$ ($1.13$) for the main panel (inset) in (b). 
    }
\end{figure*}

\section*{Floquet PXP model and protected nullspaces}
\label{sec1}
We consider a finite chain of even length $L$ with periodic boundary conditions (PBC) in the strong Rydberg blockade regime. In this constrained Hilbert space, where no two consecutive sites can be in a Rydberg excited state, $H(t)$ is given by 
\begin{equation}
H(t) = - w_0 \sum_{j=1}^{L} \tilde{\sigma}_j^x  \ + \  \frac{\lambda(t)}{2} \sum_{j=1}^{L} \sigma_j^z,
\label{eq1}
\end{equation}
where $\sigma_j^{ \alpha}$ for $\alpha =x,y,z$ are the Pauli matrices on site $j$ and $\tilde{\sigma}_j^{ \alpha} = P_{j-1} \, \sigma_j^{ \alpha} \ P_{j+1}$ is a projected operator with $P_k = \ket{\downarrow_k} \bra{\downarrow_k}$ as the down spin (or equivalently Rydberg ground state) projector on site $k$. We set $\hbar=1$, $w_0=1$ and scale all energies (frequencies) in units of $w_0$ ($w_0/\hbar)$. Here $\lambda (t)$ is driven periodically in time with $\lambda(t)=-\lambda$ for $0 \leq t <T/2$ and $\lambda(t)=+\lambda$ for $T/2 \leq t < T$.
The Floquet unitary equals 
\begin{equation} \label{eq3}
\begin{aligned}
    & U (T,0) = e^{-i\,T\,H_F} = e^{-i \,(T/2)\, H(\lambda)} \  \ e^{-i \,(T/2)\, H(-\lambda)} \\ & \text{with} \quad H(\lambda) = \frac{\lambda}{2} \sum_j \sigma_j^z  \ - \sum_j \tilde{\sigma}_j^x.
\end{aligned}
\end{equation}

The parity operator ${\mathcal P}$, which maps a site $i$ to $L-i+1$, is a symmetry of $U(T,0)$ along with translation by one lattice unit. We focus on the largest symmetry sector with total momentum $K=0$ and parity ${\mathcal P}=+1$. Defining the unitary chiral operator $\mathcal{C}=\prod_{j=1}^L \sigma_j^z$, it satisfies $\mathcal{C}^{-1}=\mathcal{C}$ and~\cite{Mukherjee2020a} 
\begin{eqnarray}
U^{-1}(T,0) = \mathcal{C} U(T,0) \mathcal{C}.
\label{eq:indexU}
\end{eqnarray}
Eigenvalues of $U(T,0)$, which are complex numbers $\exp(i\theta)$, come in $\pm \theta$ pairs from Eq.~\ref{eq:indexU}. Only the eigenstates with $\theta=0,\pi$ contribute to $\mathrm{Tr}(\mathcal{C})$ since these modes are also eigenstates of $\mathcal{C}$ with eigenvalues $\pm 1$. Exact diagonalization (ED) calculations on finite chains show no $\pi$-modes (see~\cite{supp_mat} for details), and all the $0$-modes have $\mathcal{C}=+1$ eigenvalues across the parameter ranges and system sizes studied here. This implies that the number of zero-quasienergy modes $\mathcal{N}_0(L)$ of $U(T,0)$ equals $\mathrm{Tr}[\mathcal{C}]_{K=0,{\mathcal P}=+1}$ independent of the drive parameters for a fixed $L$. Since $U(T,0)$ has no $\pi$-modes, the principal logarithm unambiguously defines \(H_F :=(i/T)\ln U(T,0)\), which can be obtained using standard matrix logarithm routines~\cite{Virtanen_2020, doi:10.1137/110852553}. $H_F$ satisfies $\{H_F, \mathcal{C}\}=0$ and has these $\theta=0$ modes of $U(T,0)$ as exact zero modes. The dimensionality, $\mathcal{N}_0(L)$, of the nullspace of $H_F$ is protected by an index theorem~\cite{Turner2018pxp1, Turner2018pxp2, Schecter2018_index} and equals the corresponding number for the static PXP model~\cite{Buijsman2022}. The nullspace itself continues to change with the drive parameters.  

Let $\{ |Z_i \rangle \}$ be an orthonormal basis of this zero-quasienergy nullspace. For any normalized state $|\psi \rangle$, we define 
\begin{eqnarray}
    W_0(\psi) = \langle \psi | \mathbb{P}_0 |\psi \rangle = \sum_i |\langle Z_i |\psi \rangle|^2, 
    \label{eq:defW0}
\end{eqnarray}
and the normalized projected descendant $|Z_\psi \rangle$ of the parent $|\psi \rangle$ equals~\cite{supp_mat} 
\begin{eqnarray}
|Z_\psi \rangle =\frac{\mathbb{P}_0 |\psi \rangle}{\sqrt{W_0(\psi)}} = \sqrt{W_0(\psi)} |\psi\rangle + \sqrt{1-W_0(\psi)}|\psi_{\perp} \rangle
\label{eq:defdescendant}
\end{eqnarray}
with $\langle \psi | \psi_{\perp} \rangle=0$. A random parent state with a well-defined $\mathcal{C}=+1$ has mean nullspace weight of 
\begin{eqnarray}
    W_0^{\mathrm{typ}}(L) = \frac{2\mathcal{N}_0(L)}{\mathcal{D}_L+ \mathcal{N}_0(L)} \sim e^{-\gamma_{\mathrm{typ}}L} \ ,
    \label{eq:W0typ}
\end{eqnarray}
where $\mathcal{D}_L$ is the dimension of the $(K=0, \mathcal{P}=+1)$ sector for $L$ spins. This follows from the dimension of the $\mathcal{C}=+1$ block within $K=0,\mathcal{P}=+1$ sector being $(\mathcal{D}_L+\mathcal{N}_0(L))/2$. In an extended many-body system, \(O(1)\) dressing per finite spatial region generically accumulates extensively, so that the global fidelity between a parent and its dressed descendant may naturally decay as \(W_0(\psi)\sim e^{-\gamma_\psi L}\)~\cite{supp_mat}. Indeed, the corresponding typical weight $W_0^{\mathrm{typ}}(L)$ is exponentially small in $L$ from Hilbert-space counting alone (Eq.~\ref{eq:W0typ}). The relevant signature of an anomalous descendant is therefore not a non-vanishing \(W_0 (\psi)\) for $L \gg 1$, but a parametrically smaller decay rate, \(\gamma_\psi<\gamma_{\rm typ}\) or equivalently, 
\begin{eqnarray}
    W_0(\psi)/W_0^{\mathrm{typ}} \sim e^{\gamma_e L}, \gamma_e=\gamma_{\mathrm{typ}}-\gamma_{\psi}>0.
    \label{eq:expenh}
\end{eqnarray}

To understand why particular parents have anomalously enhanced nullspace
weight, we use Floquet perturbation theory (FPT)~\cite{Mukherjee2020b}, assuming $\lambda \gg 1$, which organizes $H_F=H_F^{(1)}+H_F^{(3)}+\cdots$~\cite{supp_mat}:
\begin{eqnarray}
    H_F^{(1)} &=& -\frac{\sin(\pi x)}{\pi x} \sum_j \left(e^{-i \pi x} \tilde{\sigma}_j^{+}+\mathrm{h.c.}\right) \nonumber \\
    H_F^{(3)} &=& \sum_j ( A_0 P_{j-2}\sigma_{j-1}^{+} \sigma_j^{-} \sigma_{j+1}^{+}P_{j+2} \nonumber \\
    &-& A_0 \tilde{\sigma}^{+}_jP_{j+2} -A_0 P_{j-2}\tilde{\sigma}_j^{+}-A_0 \tilde{\sigma}^{+}_j  + \mathrm{h.c.})
    \label{FPT}
\end{eqnarray}
where $\sigma_j^{\pm}=(\sigma_j^x \pm i \sigma_j^y)/2$, $A_0 = \frac{e^{-i 4 \pi x}}{24 \pi i \omega_D^2 x^3 } [e^{6i \pi x} + 3 e^{2i \pi x} (1 + 4i \pi x)   + 2 (1 - 3e^{4i \pi x})]$ with $\omega_D = 2 \pi /T$ being the drive frequency and $x := \lambda/(2 \omega_D)$. Note that FPT gives only terms that connect Fock states (in the $\sigma^z$ basis) that differ by an odd number of up spins at all orders, consistent with $\{ H_F, \mathcal{C}\}=0$. 
We will consider
this perturbative $H_F$ (Eq.~\ref{FPT}) to analytically understand
the two zero-quasienergy Floquet scars in the limit where $W_0(\psi) \rightarrow 1$. 

\begin{figure*} \label{fig3_PRL_resub}
     \includegraphics[width=0.48\linewidth]{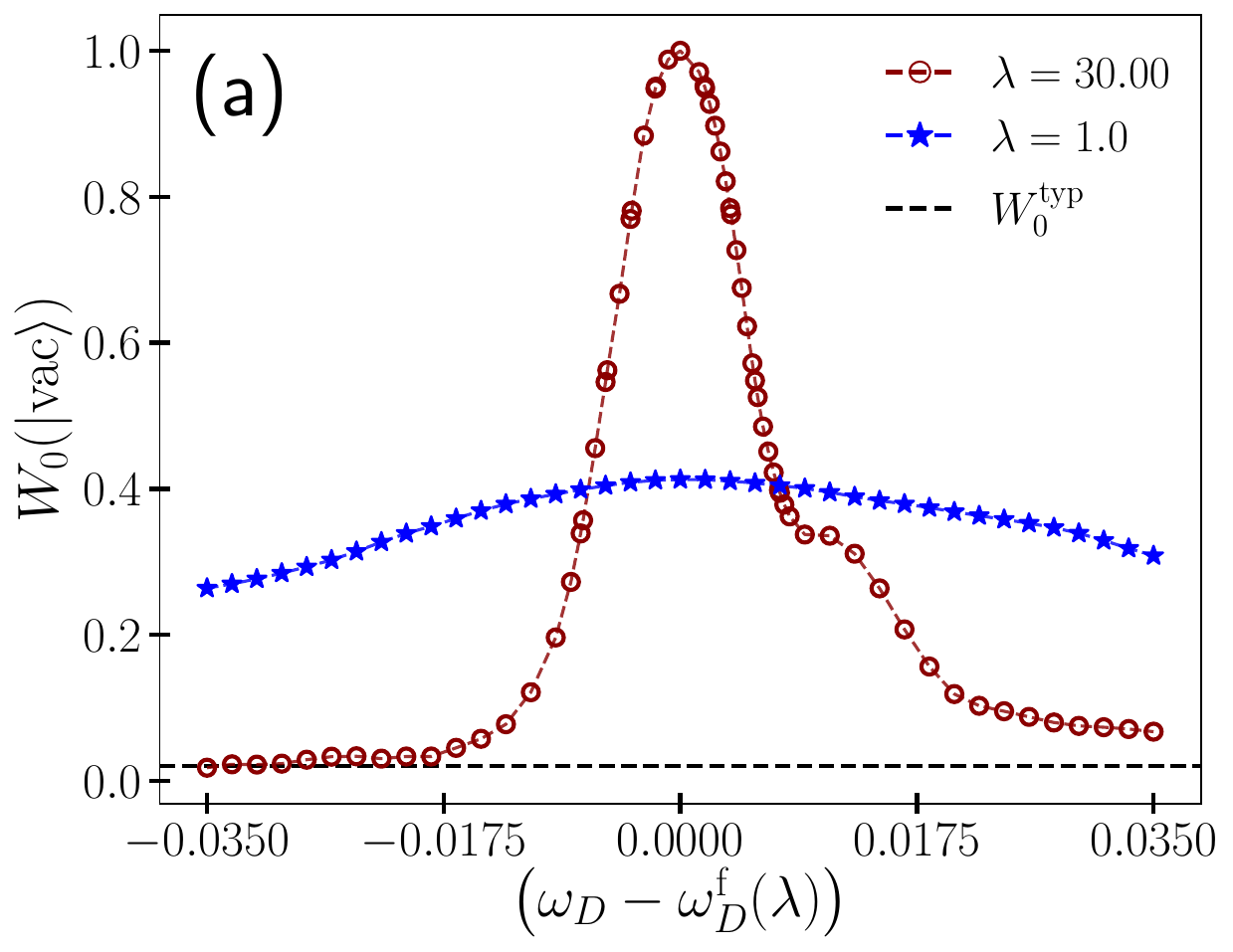}
     \includegraphics[width=0.48\linewidth]{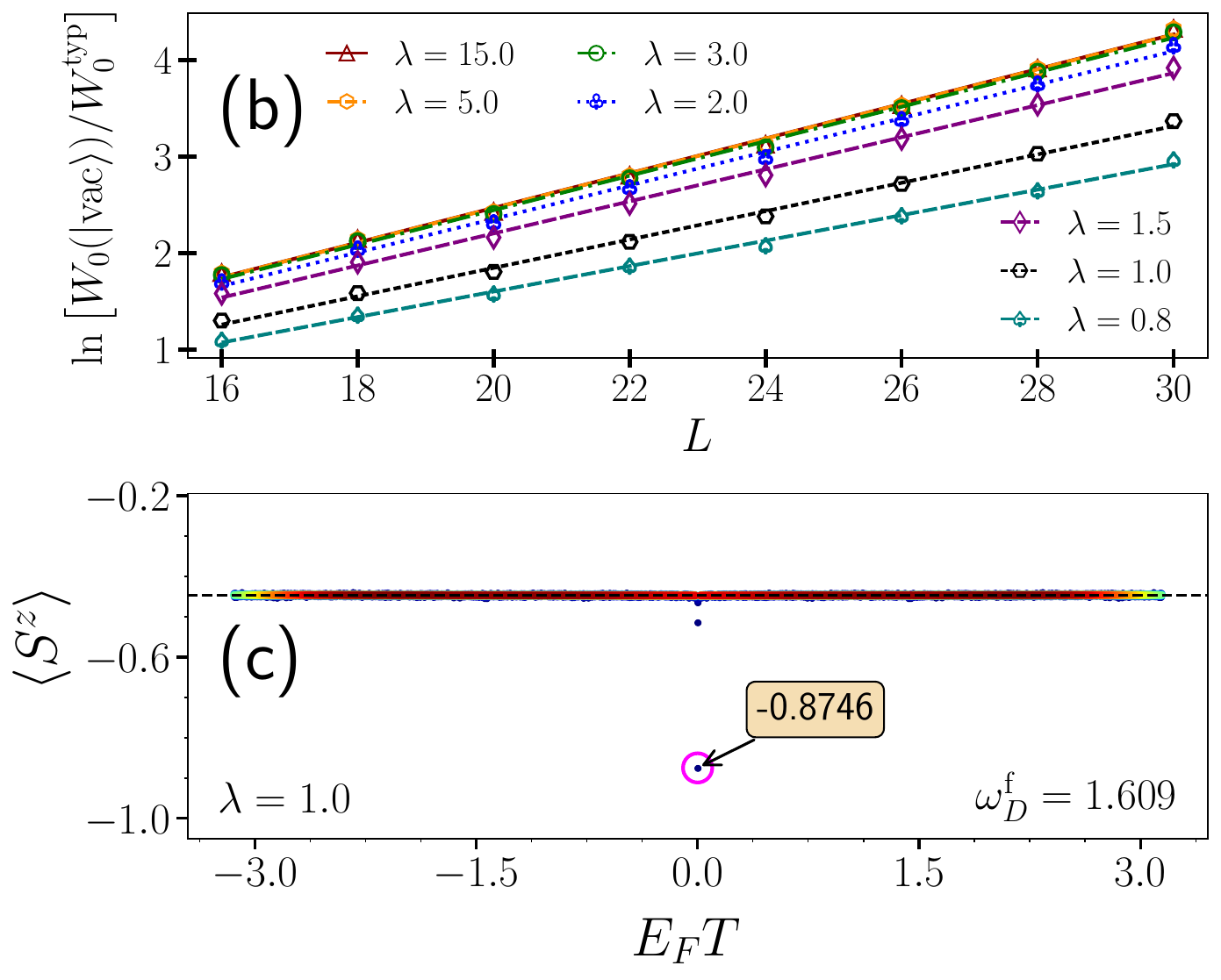}
     \caption{(a) $W_0(\ket{\mathrm{vac}})$ for $\lambda=30$ (red open circles) and $\lambda=1$ (blue filled asterisks) around $\omega_D^{\mathrm{f}} = 15.09954,  1.609$ respectively, where \(W_0\) attains its maximum for the respective values of \(\lambda\), for the system size $L=28$. $W_0^{\rm typ}\approx 0.02$ for $L=28$ and is shown by the horizontal black dashed line. (b) $\ln \left[ W_0(\ket{\rm vac})/ W_0^{\rm typ}\right]$ as a function of $L$ using $16 \leq L \leq 30$ for different $\lambda$ where $\omega_D=\omega^{\mathrm{f}}_D (\lambda)$. (c) Plot of $\langle S^z \rangle := \frac{1}{L} \sum_j \langle \sigma_j^z\rangle $ for eigenstates of $H_F$, where $E_F T \in [-\pi,\pi)$, at $\lambda=1, \, \omega_D^{\mathrm{f}} =1.609$ for $L=30$ ($K=0,\mathcal{P}=+1$ sector).  The dressed vacuum scar is circled in magenta, and its $\langle S^z \rangle$ is displayed. The density of states is indicated by a color map where warmer colors correspond to higher density of states. The infinite-temperature ensemble average ($\approx -1/\sqrt{5}$) is shown as a horizontal black dashed line. 
     }
\end{figure*}

\section*{Dressed Ivanov-Motrunich scar}
\label{sec2}

Recent work~\cite{Ivanov2025} found an exact zero mode of the PXP model that can be expressed as follows:
\begin{eqnarray}
\ket{\Lambda} := \frac{1}{\sqrt{\mathcal{D}_{\frac{L}{2}}}} \sum_{\ket{f} \in \mathcal{H}^{\mathrm{PBC}}_{L/2}} (-1)^{w_f}  \ket{f}_{1,\cdots,\frac{L}{2}} \otimes \ket{f}_{\frac{L}{2}+1,\cdots,L} \label{eqIMstate}
\end{eqnarray}
where $\mathcal{D}_{L/2}$ is the dimensionality of the Hilbert space $\mathcal{H}^{\mathrm{PBC}}_{L/2}$ of $L/2$ spins under PBC. The sum is over all Fock states $\ket{f}$ (in the computational basis) in $\mathcal{H}_{L/2}$, and $w_f$ equals the number of up spins in $\ket{f}$. This state is highly entangled, since all antipodal spins at sites $i$ and $L/2+i$ are perfectly correlated (Fig.~\ref{fig_1_PRL_resub}). By construction, $\ket{\Lambda}$ has quantum numbers $K=0, \mathcal{P}=+1$. 

In fact, $|\Lambda \rangle$ was shown to be an exact zero mode of a whole class of Hamiltonians $H$ that satisfy (i) $\langle g|H|f\rangle = \langle f|H|g \rangle$ and (ii) $(-1)^{w_f}=-(-1)^{w_g}$ whenever the matrix elements of $H$ are nonzero, without any assumption of locality. Although condition (ii) is automatically satisfied by $H_F$ for its nonzero matrix elements given that $\{H_F, \mathcal{C}\}=0$, condition (i) is not. 

For a generic $x:=\lambda/(2\omega_D)$, $H_F^{(1)}$ is nonzero and complex Hermitian in the computational basis. However, a local unitary rotation $\mathcal{U}(x)=\exp(-i\pi x\sum_j \sigma_j^z/2)$ transforms $H_F^{(1)}$ to $\mathcal{U}(x)^\dagger H_F^{(1)} \mathcal{U}(x) = -\sin(\pi x)/(\pi x) \sum_j \tilde{\sigma}_j^x$ which is real symmetric. Thus,
\begin{eqnarray}
    |\Lambda_r \rangle =  \mathcal{U}(x) \ket{\Lambda}=\exp \left(-i\frac{\pi x}{2} \sum_j \sigma_j^z \right)|\Lambda \rangle
    \label{rotatedIM}
\end{eqnarray}
is a zero mode of $H_F^{(1)}$. However, $\mathcal{U}(x)$ does not simultaneously convert higher-order terms like $H_F^{(3)}$ to be real symmetric.  As $\omega_D$ decreases for a fixed $x$, these terms become increasingly important and progressively displace $|\Lambda_r \rangle$ away from the protected Floquet nullspace.

This effect can be clearly seen from ED for $L=28$ and $x=1.2$ (Fig.~\ref{fig2_PRL_resub}(a), main panel), which shows that $W_0(\ket{\Lambda_r}) \rightarrow 1$ at $\omega_D \gg 1$ followed by two plateaus separated by a broad knee. The onset of the knee at  $\omega_D \approx 7.06$ occurs as $H_F^{(3)}$ becomes important, while the eventual plateau where $W_0(\ket{\Lambda_r}) \approx W_0^{\mathrm{typ}}$ below $\omega_D \approx 1.48$ requires still higher, increasing nonlocal terms. Crucially, the absolute decrease in $W_0$ is not, by itself, the scar criterion. The finite-size scaling of $\ln[W_0(\ket{\Lambda_r})/W_0^{\mathrm{typ}}]$ shows growth with $L$ over a broad range above the low-frequency plateau, consistent with a $\gamma_e>0$ (Eq.~\ref{eq:expenh}) over the accessible sizes $16 \leq L \leq 30$ (Fig.~\ref{fig2_PRL_resub}(a), inset).

Projection onto the protected Floquet nullspace can preserve characteristic parent correlations even after substantial loss of global fidelity. At \(x=1.2\) and \(\omega_D=1.75\), for example, the antipodal correlator, $O_{i,i}=\sigma_i^z\sigma_{i+L/2}^z$, of \(|Z_{\Lambda_r}\rangle\) (Eq.~\ref{eq:defdescendant}) remains strongly enhanced [Fig.~\ref{fig2_PRL_resub}(b), main panel], despite \(W_0\simeq0.0918\) at \(L=30\), compared with \(W_0^{\rm typ}\simeq0.0132\). This contrasts sharply with the broad distribution of typical Floquet eigenstates centered near \(\langle \sigma^z\rangle^2=1/5\). On the other hand, at \(\omega_D=1.0\), where \(W_0\simeq0.0149\approx W_0^{\rm typ}\), the same correlator moves much closer to \(1/5\) [Fig.~\ref{fig2_PRL_resub}(b), inset], signifying the loss of parent-like correlations upon projection.

\section*{Dressed vacuum scar}
\label{sec4}
We next consider the Rydberg vacuum $\ket{\mathrm{vac}} = \ket{\downarrow\cdots \downarrow \downarrow \downarrow }$ (Fig.~\ref{fig_1_PRL_resub}). 
When $\lambda \gg 1$, third-order FPT gives 
\begin{eqnarray}
    (H_F^{(1)}+H_F^{(3)})|\downarrow\cdots \downarrow \downarrow \downarrow \rangle = w_{\mathrm{eff}} \sum_{j=0}^{L-1} T^j |\downarrow\cdots \downarrow \downarrow \uparrow \rangle
    \label{Honvac}
\end{eqnarray}
where $T^r$ represents a translation of $r$ lattice sites and $w_{\mathrm{eff}} = w^{(1)}+w^{(3)}$, 
with $w^{(1)}=-\exp(-i \pi x) \sin(\pi x)/(\pi x)$ $ \Big(w^{(3)}=-3A_0 \Big)$ generated by $H_F^{(1)}$ $ \Big(H_F^{(3)} \Big)$. Near the dynamical freezing frequencies identified in Ref.~\cite{Mukherjee2020b}, $|w_{\mathrm{eff}}|$ becomes anomalously small, making the vacuum a natural parent with a strongly enhanced projection onto the protected Floquet nullspace. Importantly, $w_{\mathrm{eff}}$ receives contributions from all orders in FPT. ED at $\lambda=30$ already shows a large nonperturbative renormalization of $|w_{\mathrm{eff}}|$ near the highest freezing frequency~\cite{supp_mat} (which we denote by $\omega_D^{\mathrm{f}}(\lambda)$), motivating us to study its anomalous nullspace weight beyond the perturbative regime.

The distinction between the perturbative freezing and the more general nullspace-projection mechanism is illustrated in Fig.~\ref{fig3_PRL_resub} (a) using data for $L=28$. For $\lambda=30$, $W_0(\ket{\mathrm{vac}})$ displays a narrow peak reaching near $1$ around $\omega_D^f$, consistent with the near-perfect freezing discussed in Ref.~\cite{Mukherjee2020b}. By contrast, for $\lambda=1$, where low-order FPT is no longer a controlled approximation, the peak is much broader, and the maximum is substantially reduced to $W_0 \sim 0.4$. Importantly, the value remains far above $W_0^{\mathrm{typ}} \approx 0.02$ for $L=28$ over the entire range of $\omega_D$ in Fig.~\ref{fig3_PRL_resub}(a). Thus, the dressed scar $|Z_{\mathrm{vac}} \rangle$ (Eq.~\ref{eq:defdescendant}) persists even after the near-perfect freezing regime has disappeared.

Fig.~\ref{fig3_PRL_resub}(b) shows $\ln[W_0(\ket{\mathrm{vac}})/W_0^{\mathrm{typ}}]$ for $16 \leq L \leq 30$ evaluated at $\omega_D =\omega_D^f(\lambda)$. For all $0.8 \leq \lambda \leq 15.0$, this quantity increases with $L$ over the accessible system sizes, consistent with Eq.~\ref{eq:expenh} with a positive $\gamma_e$. Over the accessible system sizes, the projected vacuum descendant becomes increasingly atypical relative to a random state in the same symmetry sector, even when its absolute parent-state overlap $W_0$ decreases with system size~\cite{supp_mat}.

The anomalous nature of $|Z_{\mathrm{vac}}\rangle$ (Eq.~\ref{eq:defdescendant}) is reflected in the polarization even in the nonperturbative regime of $\lambda \sim O(1)$. To illustrate this, we plot $\langle S^z \rangle = \langle \sum_{i=1}^L \sigma_i^z \rangle /L$ for $L=30$ at $\lambda=1$ and $\omega_D^f=1.609$ for each eigenstate of $U(T,0)$ (Eq.~\ref{eq3}) using ED as a function of the corresponding quasienergy $E_F T$ (defined in the first Floquet Brillouin zone) (Fig.~\ref{fig3_PRL_resub} (c)). While the nonzero-quasienergy eigenstates cluster close to the infinite-temperature value of the polarization, consistent with Floquet ETH, $|Z_{\mathrm{vac}}\rangle$ remains strongly polarized with $\langle S^z \rangle =-0.8746$ (Fig.~\ref{fig3_PRL_resub} (c)). The dressed vacuum scar therefore remains a conspicuous outlier even when $H_F$ is strongly nonlocal. 
The value of $\langle S^z\rangle \simeq -0.87$ (Fig.\ \ref{fig3_PRL_resub} (c)) highlights the importance of the terms represented by $|\psi_{\perp} \rangle$ in Eq.\ \ref{eq:defdescendant}; without them, $\langle S^z \rangle = -W_0 \simeq -0.38$ acquires a substantially smaller value. 
Thus the orthogonal dressing component itself retains substantial parent-specific structure since $\langle \psi_\perp|S^z|\psi_\perp \rangle \simeq -0.81$.

\section*{Conclusions and outlook}
We have shown that an exponentially large (in system size) symmetry-protected Floquet nullspace can host anomalous zero-quasienergy states selected by their relation to physically distinguished parent states. For a parent \(|\psi\rangle\), projection onto the protected nullspace defines an exact descendant \(|Z_\psi\rangle\). The nontrivial content of this construction is not the projection itself, but the existence of parents whose nullspace weight is parametrically enhanced relative to the typical value. Our finite-size results are consistent with \(W_0(\psi)/W_0^{\rm typ}\) growing exponentially with system size, even when the absolute parent-state overlap decreases exponentially. Thus global fidelity with the parent need not remain finite for a Floquet scar to remain anomalous.

We demonstrated this mechanism for two qualitatively different parents in the periodically driven PXP chain. A locally rotated, volume-entangled Ivanov–Motrunich state gives rise to a zero-quasienergy descendant that retains its characteristic long-range antipodal correlations as long as its projection weight remains well above the typical nullspace scale. The unentangled Rydberg vacuum provides a complementary example: its descendant evolves continuously from the near-perfect dynamical-freezing regime at large drive amplitude to a strongly dressed scar that remains a conspicuous polarization outlier even when low-order Floquet perturbation theory has broken down and \(H_F\) is strongly nonlocal. Complementing these eigenstate signatures, long-time stroboscopic dynamics from both parent states retain the corresponding anomalous correlations over the same parameter regimes (see End Matter).

These results suggest that symmetry-protected Floquet nullspaces provide a more general setting in which special many-body structures can survive projection despite extensive loss of global fidelity. It will be interesting to determine how generally such anomalously enhanced descendants occur in interacting Floquet systems. 

\section*{Acknowledgments} KS thanks DST, India, for support through the project JCB/2021/000030. SR and AS thank Diptiman Sen for discussions. SR acknowledges the support from UGC under the JRF Scheme and thanks Sinchan Ghosh for discussions. B.~M.\@ acknowledges funding from DST, Government of India, via the INSPIRE Faculty program. The authors used ChatGPT (OpenAI), GPT-5.6 Sol, to help refine the presentation and articulation of scientific arguments. The authors critically evaluated and independently verified all AI-assisted output and take full responsibility for the scientific content, calculations, results, interpretations, and conclusions.

\bibliography{citations}

\appendix*

\section* {{\bf{End Matter}}}
\begin{figure} \label{figEM1}
    \includegraphics[width=\linewidth]{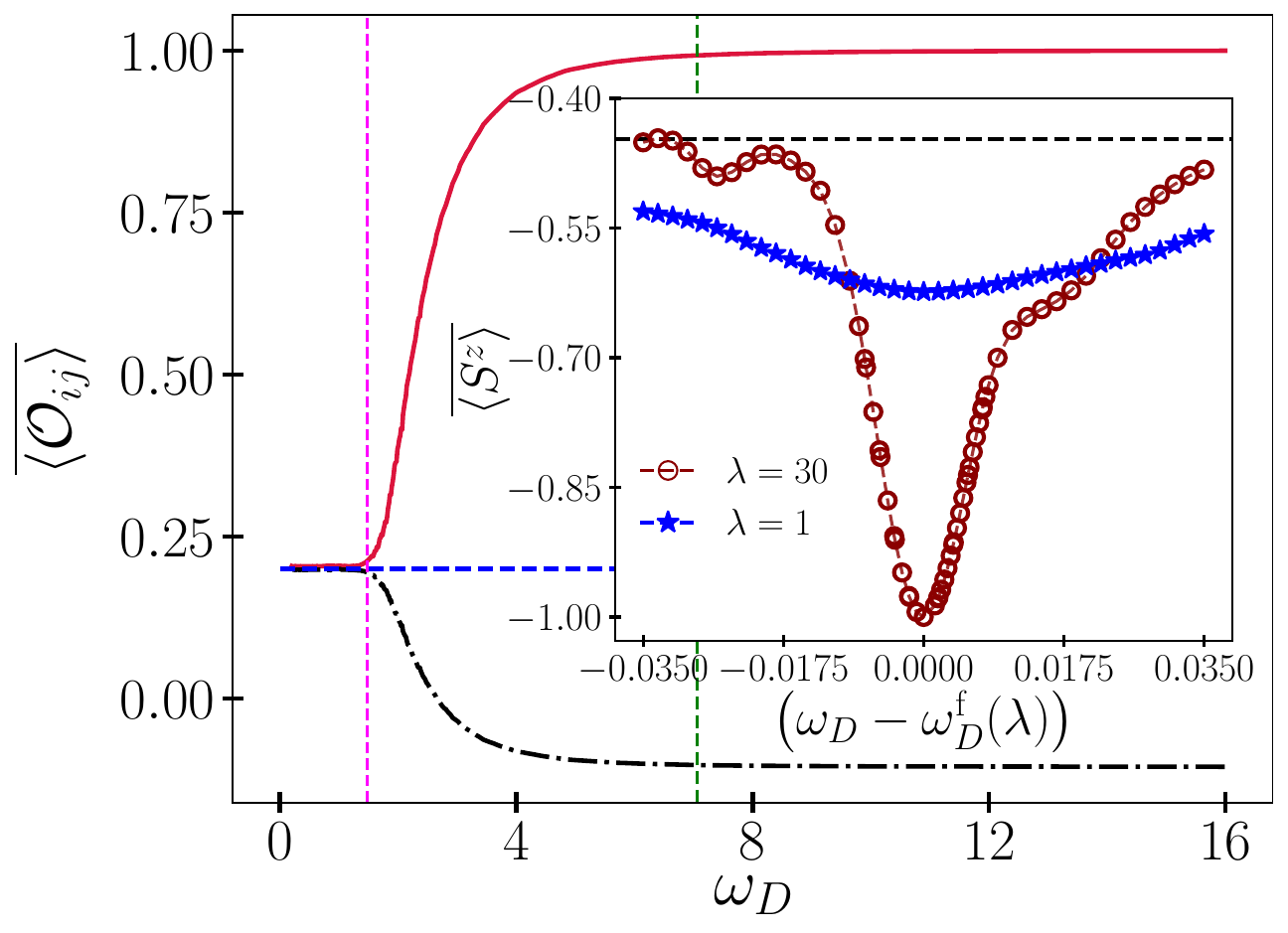}
    \caption{Time-averaged spin-spin correlation $\overline{\langle \mathcal{O}_{ij} \rangle} = \overline{\langle \sigma_i^z \sigma_{L/2 +j}^z \rangle}$ starting from $\ket{\Lambda_r}$ for $j=i$ (red) and $j=i+1$ (black) for the system size $L=28$ at $x=1.2$. A horizontal blue dotted line at $0.2$ serves as a visual guide for the corresponding infinite-temperature ensemble value when $L \gg 1$. The vertical dotted line in green (magenta) corresponds to the onset of the first (second) plateau at $\omega_D \approx 7.06$ ($\omega_D \approx 1.48$) in Fig.~\ref{fig2_PRL_resub}(a). Inset shows time-averaged magnetization $\overline{\langle S^z} \rangle = \frac{1}{L} \overline{ \langle \sum_{j=1}^L \sigma_j^z\rangle}$ starting from vacuum state near $\omega_D^{f}$ of $\lambda=30$ and $\lambda=1$ for the same system size $L=28$. The average is taken over $n=5001$ to $n=6000$ stroboscopic cycles for each $\omega_D$ in both the main figure and the inset. The infinite-temperature ensemble average ($ -1/\sqrt{5}$ for $L \gg 1$) is shown as a horizontal black dashed line as a visual guide.}
\end{figure}
We now examine the long-time stroboscopic dynamics starting from the corresponding parent states and show that substantial memory of their characteristic correlations and polarization persists over the drive regimes where the projected descendants remain anomalous.

Starting the Floquet dynamics from an initial rotated IM parent state $|\Lambda_r \rangle$ (Eq.~\ref{rotatedIM} in the main text) at a fixed $x=1.2$, we compute a time-averaged two-point spin correlation $\overline{\langle \mathcal{O}_{i,j}\rangle} = \overline{\langle \sigma_i^z \sigma_{L/2 +j}^z \rangle}$ for $L=28$ using the same drive frequencies as displayed in Fig.~\ref{fig2_PRL_resub} (a) (main panel) in the main text. Here $i=j$ corresponds to antipodal spins, and $j=i+1$ indicates two spins closest to being an antipodal pair. The behavior of $\overline{\langle \mathcal{O}_{i,j}\rangle}$ is shown in Fig.~\ref{figEM1}, main panel. The two correlators remain clearly distinguishable down to the onset of the second plateau at \(\omega_D\simeq1.5\), demonstrating persistent memory of the parent-state correlation structure. At large $\omega_D$, since \(|\Lambda_r\rangle\) lies very close to the Floquet nullspace, the values of these two-spin correlators approach $\langle \Lambda_r | \mathcal{O}_{i,j} |\Lambda_r \rangle = \langle \Lambda | \mathcal{O}_{i,j} |\Lambda \rangle$, where the equality follows because $\mathcal{O}_{i,j}$ commutes with $\mathcal{U}(x)$ for this local operator. 

Using the form of $|\Lambda \rangle$ in Eq.~\ref{eqIMstate} (main text), we can compute the correlators straightforwardly for $i=j$ and $j=i+1$. For $i=j$, $\langle \Lambda | \mathcal{O}_{i,j} |\Lambda \rangle=1$ since the antipodal pairs are perfectly correlated (Fig.~\ref{fig_1_PRL_resub}), which is consistent with the behavior of the time-averaged $\overline{\langle \mathcal{O}_{ii} \rangle}$ at large $\omega_D$ (Fig.~\ref{figEM1}, main panel). 

A similar computation can be carried out for $\langle \Lambda | \mathcal{O}_{i,j} |\Lambda \rangle$ for $j=i+1$. From Eq.~\ref{eqIMstate}, the sites $j$ and $L/2 + j$ are identified and hence $\sigma_i^z\sigma^z_{L/2+i+1} \ket{\Lambda}=\sigma_i^z\sigma^z_{i+1} \ket{\Lambda}$. Furthermore, writing $\sigma_k^z$ as $(2n_k-1)$ with $n_k$ being the number operator at site $k$, we get $\sigma_i^z\sigma^z_{i+1} = 1 - 2 (n_i + n_{i+1})$ when the Rydberg blockade constraint $n_in_{i+1}= 0$ is implemented. Since $\ket{\Lambda}$ has support only from Fock states $\ket{f} \in \mathcal{H}_{L/2}^{\mathrm{PBC}}$, the expectation value $\bra{\Lambda}\mathcal{O}_{i, i+1}\ket{\Lambda}$ simplifies to $1- \frac{4}{\mathcal{D}_{L/2}}\, \mathrm{tr}(n_L)$, where we have replaced the trace of $n_i$ and $n_{i+1}$ with $n_L$ without the loss of generality. This trace can be calculated using a transfer matrix approach to yield
\begin{eqnarray}
\frac{1}{\mathcal{D}_{L/2}}\mathrm{tr}(n_L) = \frac{F_{L/2 -1} }{F_{L/2 -1} + F_{L/2 +1}} \rightarrow \frac{1}{1 + \phi^2}
\label{eqEM1}
\end{eqnarray}
where $F_n$ denotes the Fibonacci numbers defined by $F_n+F_{n+1}=F_{n+2}$ with $F_1=F_2=1$. The last expression in Eq.~\ref{eqEM1} is obtained taking $L \rightarrow \infty$ where $\phi=(1+\sqrt{5})/2$ equals the golden mean. Since $\phi^2 = \phi + 1$, this leads to $\bra{\Lambda}\mathcal{O}_{i, i+1}\ket{\Lambda} \approx -(1-2/\sqrt{5}) \approx -0.1056$ as $L \gg 1$, which is consistent with the behavior of the time-averaged $\langle \mathcal{O}_{ij} \rangle$ with $j=i+1$ at large $\omega_D$ (Fig.~\ref{figEM1}, main panel).

As $\omega_D$ decreases, the time-averaged values of these two correlators change since $W_0(|\Lambda_r \rangle)$ decreases from $1$. However, the memory effect in two-body correlators remains clear because these correlators stay different from each other. This difference decreases as $\omega_D$ decreases and vanishes as the second $\omega_D$ plateau is reached, beyond which the projected descendant ceases to be anomalous relative to the typical nullspace scale since $W_0 \approx W_0^{\mathrm{typ}}$ (Fig.~\ref{fig2_PRL_resub}(a), main panel). In this regime, both correlators approach the infinite-temperature prediction $\langle \sigma^z \rangle^2 = 1/5$ for $L \gg 1$, as shown in Fig.~\ref{figEM1}.

Starting the Floquet dynamics from the Rydberg vacuum state and focusing on the drive frequencies used to generate Fig.~\ref{fig3_PRL_resub}(a) (main text) for $\lambda=30$ and $\lambda=1$ respectively, we see a similar memory effect in the stroboscopic dynamics when we compute the time-averaged polarization $\overline{\langle S^z \rangle}$ (Fig.~\ref{figEM1}, inset). For \(\lambda=30\), the time-averaged polarization shows a sharp minimum near \(\langle S^z\rangle=-1\), reflecting near-perfect dynamical freezing of the vacuum in the large-amplitude regime. Furthermore, as $W_0$ approaches $W_0^{\mathrm{typ}}$ from above for $\lambda=30$ (Fig.~\ref{fig3_PRL_resub} (a)) as $\omega_D$ deviates sufficiently from $\omega_D^{\mathrm{f}} (\lambda=30)$, $\overline{\langle S^z \rangle}$ also approaches its infinite-temperature value of $-1/\sqrt{5}$. In contrast, for \(\lambda=1\) the feature is considerably broader, and the time-averaged polarization nonetheless remains well separated from its infinite-temperature value of $-1/\sqrt{5}$, despite the absence of freezing. Thus, substantial memory of the vacuum persists dynamically in the strongly dressed, nonperturbative regime, precisely where its nullspace weight remains anomalously enhanced, $W_0 \gg W_0^{\mathrm{typ}}$ (Fig.~\ref{fig3_PRL_resub}(a)). 

\end{document}


\title{Supplemental Material: Dressed Floquet scars from protected zero modes in a Rydberg chain} 
\author{Saptadip Roy\textsuperscript{1},
Bhaskar Mukherjee\textsuperscript{2,3},
K.~Sengupta\textsuperscript{1} and
Arnab Sen\textsuperscript{1}} 
\affiliation{{\bf 1} School of Physical Sciences, Indian Association for the Cultivation of Science, Kolkata 700032, India \\
{\bf 2} School of Physics, University of Hyderabad, Prof. C. R. Rao Road, Gachibowli, Hyderabad 500046, India \\
{\bf 3} S. N. Bose National Centre for Basic Sciences, Block JD, Sector III, Salt Lake, Kolkata 700106, India}

\date{\today}

\maketitle

\setcounter{section}{0}
\setcounter{equation}{0}
\setcounter{figure}{0}
\setcounter{table}{0}
\setcounter{page}{1}

\onecolumngrid

In addition to providing analytical derivations and numerical details, we examine how the reduction of parent-state fidelity accompanies the evolution of anomalous correlations, entanglement, and dynamical properties of the dressed Floquet scars.

\tableofcontents
\clearpage

\twocolumngrid

\section{Absence of $\pi$ modes in $U(T,0)$}
\label{sec_SM_H}

The absence of $\pi$ modes is important for unambiguously defining the Floquet Hamiltonian through a logarithm of the Floquet unitary $U(T,0)$. As discussed in the main text, the Floquet unitary satisfies $CU(T,0)C=U^{-1}(T,0)$, which ensures that for each eigenvalue $\exp(i \theta)$ of $U(T,0)$, there exists another eigenvalue at $\exp(-i\theta)$. Both $\theta=0$ and $\theta=\pi$ are fixed points of this reflection symmetry between $\theta$ and $-\theta$, and hence both $\theta=0$ and $\theta=\pi$ can have protected modes. 

\begin{widetext}
\begin{figure*}[t]
    \label{figSM7}
     \includegraphics[width=0.7\linewidth]{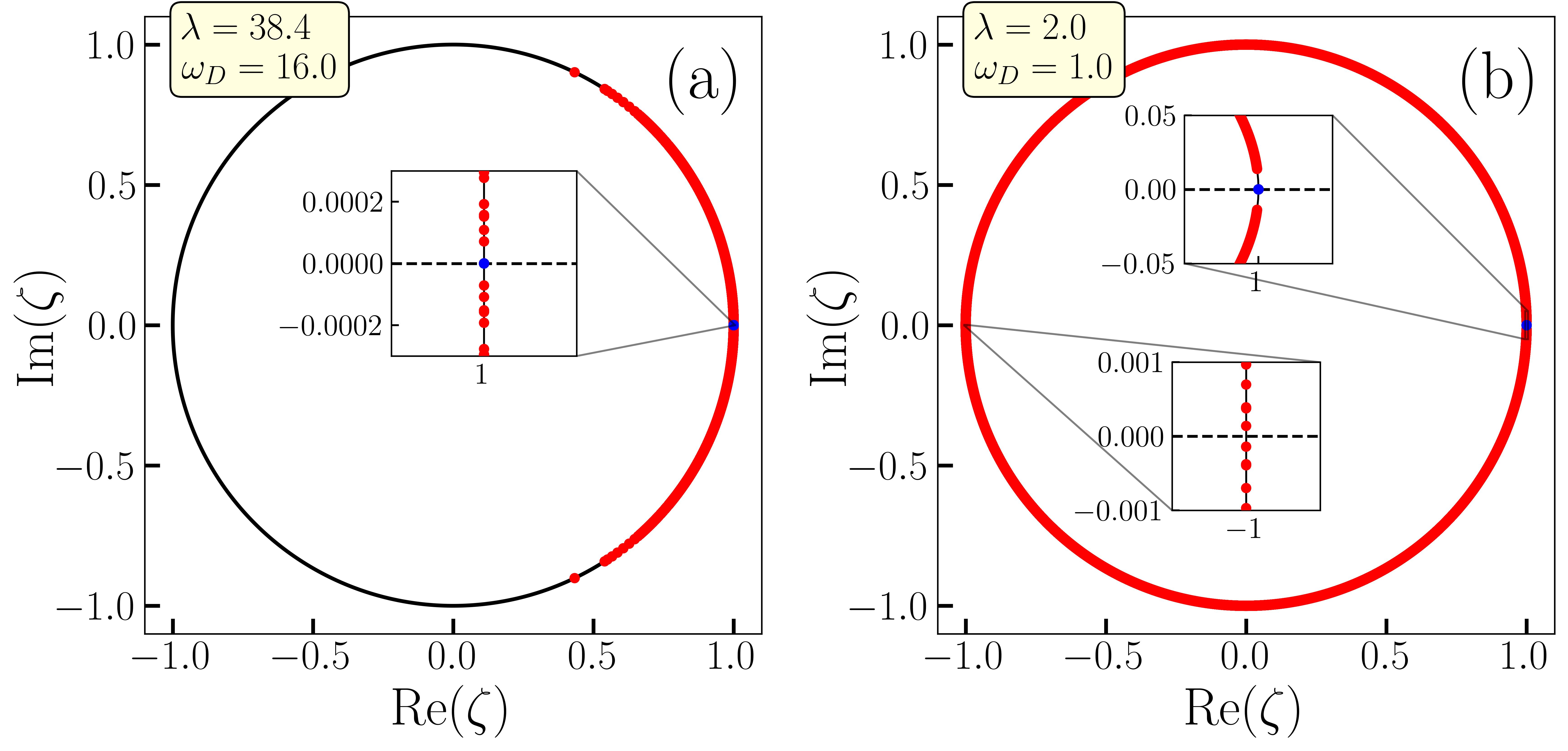}
     \caption{Plot of eigenvalues $\{\zeta = e^{i \theta} \}$ of $U(T,0)$ on a unit circle where $\theta \in (-\pi, \pi)$ for $L=30$ in the $K=0,\mathcal{P}=+1$ sector with the parameters (a) $\lambda=38.4, \, \omega_D=16$ and (b) $\lambda=2, \, \omega_D=1$ describing perturbative and non-perturbative regimes respectively. The sector dimension is $31836$. The eigenvalues 
     at $\theta=0$ ($\zeta=1$), which are $212$ in total for both cases, are highlighted in blue. The insets in both (a) and (b) have a dotted horizontal line at $\mathrm{Im}(\zeta)=0$ to visually demonstrate the presence (absence) of $\theta=0$ ($\theta=\pi$) modes.}
\end{figure*}
\end{widetext}

In Fig.~\ref{figSM7}, we plot the eigenvalues $\{\zeta=e^{i\theta}\}$ of $U(T,0)$ on the complex unit circle for $L=30$ in the $K=0,\mathcal{P}=+1$ sector. We show two representative parameter values: $(\lambda,\omega_D)=(38.4,16)$, corresponding to the perturbative regime, and $(\lambda,\omega_D)=(2,1)$, corresponding to the non-perturbative regime. In both cases, we see eigenvalues at $\theta=0$ ($\zeta=1$) which are highlighted in blue in Fig.~\ref{figSM7}, while no eigenvalue occurs for $\theta=\pi$ ($\zeta=-1$).

The absence of eigenvalues at $\zeta=-1$ ensures that the principal matrix logarithm of $U(T,0)$ is well defined without encountering a branch cut at $\theta=\pi$. Consequently, in the parameter and system-size ranges studied here, we can unambiguously define and hence numerically calculate
$H_F=\frac{i}{T}\ln U(T,0)$, with quasienergies chosen on the principal branch $\theta\in(-\pi,\pi)$. The unitary symmetry relation then translates consistently to $\{H_F,C\}=0$. Thus, within the explicitly studied regime, the $\theta=0$ states identified in Fig.~\ref{figSM7} are exact zero modes of the corresponding Floquet Hamiltonian, while no $\pi$-quasienergy modes occur. As discussed in the main text, the number of $\theta=0$ modes in the $(K=0,\mathcal{P}=+1)$ sector for a fixed $L$ remains unchanged with the drive parameters. For example, this number equals $212$ for both cases displayed in Fig.~\ref{figSM7} for $L=30$, independent of the drive parameters. We emphasize that we establish the absence of $\pi$ modes numerically over the finite system sizes and drive parameters investigated here, rather than guaranteeing it by chiral symmetry alone.  

\section{Zero modes of $H_F$ with maximum overlap with parent states}
\label{sec_SM_A}

We start by constructing the stroboscopic Floquet unitary $U(T,0)$, defined as 
\begin{equation}
\begin{aligned}
    & U (T,0) = e^{-i \,(T/2)\, H(\lambda)} \  \ e^{-i \,(T/2)\, H(-\lambda)}, \\ & \text{with} \quad H(\lambda) = \frac{\lambda}{2} \sum_j \sigma_j^z  \ - \sum_j \tilde{\sigma}_j^x,
\end{aligned}
\end{equation}
in the computational basis (or symmetry resolved basis of $K=0$, $\mathcal{P}=+1$, if required) spanning a Hilbert space $\mathcal{H}$ of dimension $\mathcal{D}$. We then diagonalize the resulting complex finite-dimensional matrix $U(T,0)$ using the standard LAPACK library. We use Schur decomposition $U = Q \, T \, Q^{\dagger}$, where $Q$ is a unitary matrix. The columns of the matrix $Q$ are orthonormal eigenvectors of $U$. Here $T$ is a diagonal matrix containing all eigenvalues $\Lambda = \{e^{i \theta}\}$ of $U$ on the unit circle in the complex plane. $U(T,0)$ satisfies $ \mathcal{C} U \mathcal{C}^{^{\dagger}} = U^{-1}$ where $\mathcal{C}^2 = \mathbbm{1}$ and has a degenerate subspace with $\theta=0$ mod $2\pi$ even after resolving all the commuting global symmetries (see main text). ED thus yields an orthonormal basis spanning the $m$-dimensional degenerate subspace $\mathcal{V}$, where $m < \mathcal{D}$, and any vector in this subspace has $\theta=0$ (mod $2\pi$). We now outline a procedure to find the eigenvector in $\mathcal{V}$ that has the maximum overlap with a given parent state $\ket{\psi} \in \mathcal{H}$ to construct the corresponding $\ket{Z_\psi}$ as defined in the main text. 

To proceed, we start with a $\mathcal{D} \times m$ rectangular matrix $V = \left \{ \ket{v_1} , \ket{v_2}, \cdots , \ket{v_m} \right \}$, where $\ket{v_i} \in \mathcal{V} \subset \mathcal{H}$ is a vector having $\mathcal{D}$ complex elements when expressed in the computational basis and $\bra{v_i}\ket{v_j}= \delta_{ij}$. Let $P=VV^{\dagger}$ be the projection operator onto the subspace $\mathcal{V}$. The projected state $\ket{\psi_p} = P \ket{\psi}$ then has non-zero norm 
\begin{align}
\|\ket{\psi_p}\| = \sqrt{W_0} = \left(\sum_{k=1}^{m} |\bra{v_k} \psi \rangle|^2\right)^{1/2}.
\end{align}
Our goal is to construct a vector in $\mathcal{V}$
\begin{equation} \label{eqZpsi}
\begin{aligned}
\ket{Z_{\psi}} &= \sum_{k=1}^{m} c_k \ket{v_k}
\end{aligned}
\end{equation}
such that $\left|\bra{Z_\psi}\psi\rangle\right|^2$ equals $W_0$ and then to form an orthonormal set of $(m-1)$ vectors $\{\ket{\phi_1}, \ket{\phi_2}, \cdots, \ket{\phi_{m-1}}\}$ each of which have the same $U$ eigenvalue. These vectors are orthogonal to $\ket{Z_{\psi}}$ and hence, by definition, to $\ket{\psi}$ since the projection of $\ket{\psi}$ onto the subspace is absorbed within $\ket{Z_{\psi}}$. Once we obtain $\ket{Z_{\psi}}$, we can satisfy the latter requirement using Gram–Schmidt orthogonalization.

\begin{figure} \label{figSM1}
    \includegraphics[width=\linewidth]{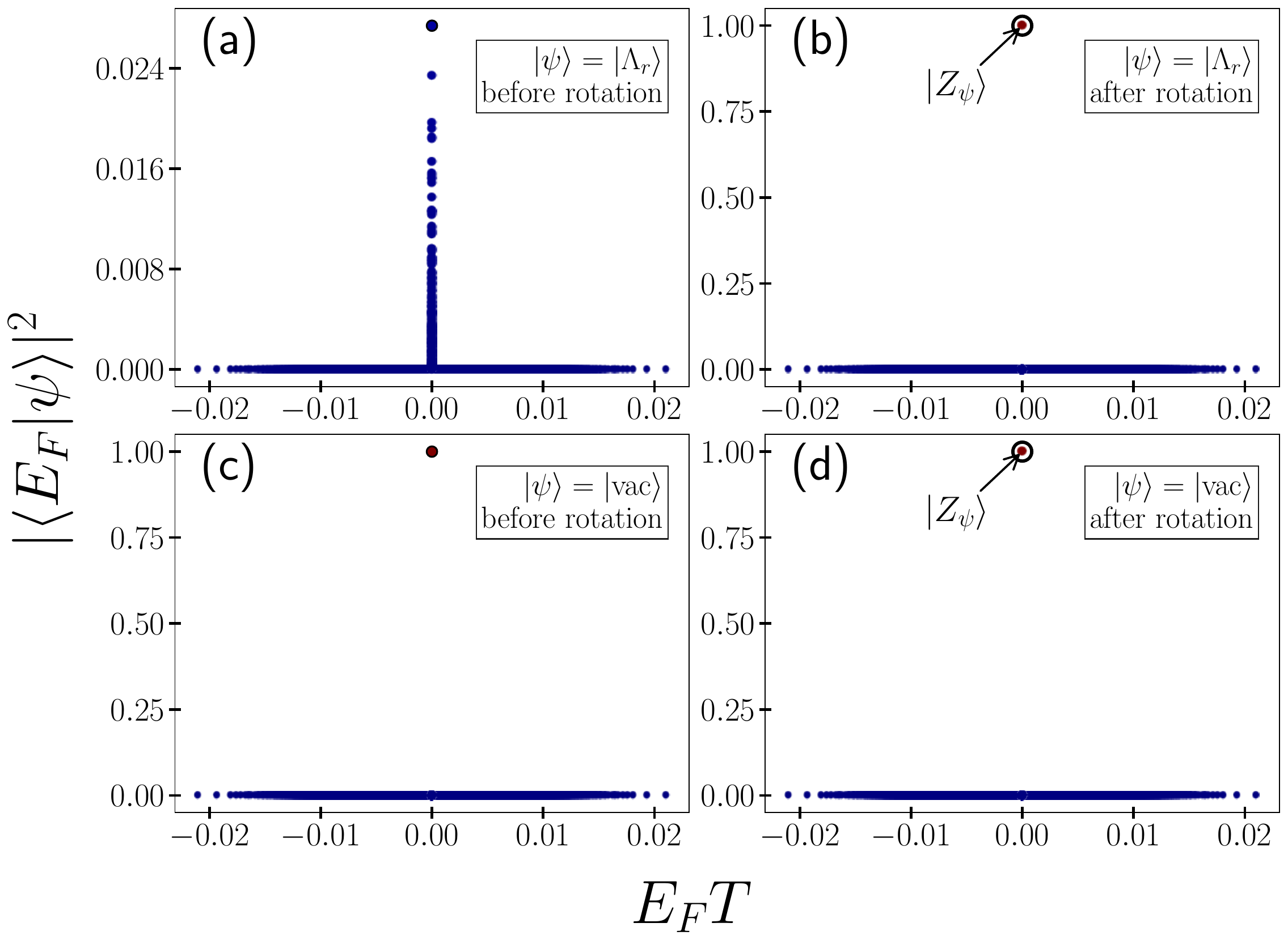} 
    \caption{Overlap of $\ket{\Lambda_r}$ ((a) and (b)) and the Rydberg vacuum state ((c) and (d)) with Floquet eigenstates $\{\ket{E_F}\}$ for $\lambda= 30, \omega_D= \omega_D^{\mathrm{f}} = 15.09954$ for $L=30$ in the $K=0, \mathcal{P}=+1$ sector. The top-left and bottom-left (right) panels show the overlaps before (after) an appropriate unitary rotation in the zero-mode subspace. After the unitary rotation, we highlight the overlap with the parent state in both cases. 
    }
\end{figure}

We construct $B = V^{\dagger} \ket{\psi}$ as an $m \times 1$ column vector and then define 
\begin{align}
C &:=
\begin{pmatrix}
c_1 \\c_2 \\\vdots \\ c_m
\end{pmatrix}
\end{align}
as an $m \times 1$ column vector that contains complex elements $c_k = \bra{v_k} Z_{\psi}\rangle$. Note that the elements of $B^{\dagger}$ are given by $\bra{\psi}v_j\rangle$ so that we have
\begin{equation} \label{eqpsi_Z_psi}
\begin{aligned}
\bra{\psi}Z_{\psi}\rangle &= \sum_k \bra{\psi}v_k\rangle \, c_k = \sum_k \, B_k^{\dagger} \,C_k \\
&= B^{\dagger} C  \ .
\end{aligned}
\end{equation}
The Cauchy–Schwarz inequality then gives
\begin{equation} \label{eqSch_inq}
\begin{aligned}
|B^{\dagger} C| & \leq \ |B| \  |C|, 
\end{aligned}
\end{equation}
while the normalization of $\ket{Z_\psi}$ \eqref{eqZpsi} implies 
\begin{equation} \label{eqnorm_Z_psi}
\begin{aligned}
|C| &= \left( \sum_k \, |c_k|^2 \right)^{1/2} = 1 \ .
\end{aligned}
\end{equation}
This guarantees that the maximum possible magnitude of the inner product in \eqref{eqpsi_Z_psi} is given by 
\begin{equation} \label{eqmax_psi_Z_psi}
\begin{aligned}
|\bra{\psi}Z_{\psi}\rangle|_{\mathrm{max}} &= |B| \ .
\end{aligned}
\end{equation}
Writing \eqref{eqZpsi} as $\ket{Z_{\psi}} = V C$, we see that the upper bound of \eqref{eqSch_inq} can be achieved by taking
$C=B/|B|$ that satisfies normalization as well as \eqref{eqmax_psi_Z_psi}. Thus, the required state $|Z_{\psi}\rangle$ in the direction of the projected parent state $\ket{\psi}$ is given by
\begin{equation} \label{eqZ_qpsi_required}
\begin{aligned}
\ket{Z_{\psi}} &= \frac{V B}{|B|} \ .
\end{aligned}
\end{equation}
Since $P^2=P=P^{\dagger}$, $|B|^2 = \bra{\psi}V V^{\dagger}\ket{\psi} = \|P\ket{\psi}\|^2$ and $VB = P \ket{\psi}$, this allows us to rewrite Eq \eqref{eqZ_qpsi_required} as 
\begin{align}
    \ket{Z_{\psi}} &= \frac{P \ket{\psi}}{\|P \ket{\psi}\|} = \frac{\ket{\psi_p}}{\sqrt{W_0}}\ .
\end{align}
Since $P\ket{\psi} = \|P\ket{\psi}\|\,\ket{Z_\psi}$, for any state $\ket{\phi}\in\mathcal{V}$ we have
\begin{align}
\langle \phi|\psi\rangle = \langle \phi|P|\psi\rangle
= \|P\ket{\psi}\|\,\langle \phi|Z_\psi\rangle .
\end{align}
Therefore, every state in $\mathcal{V}$ orthogonal to $\ket{Z_\psi}$ is automatically orthogonal to $\ket{\psi}$, as argued earlier. The procedure amounts to a unitary rotation within the $m$-dimensional degenerate eigenspace $\mathcal{V}$ which we started with. This yields an orthonormal basis in which one vector, $\ket{Z_{\psi}}$, maximizes its overlap with the parent state $\ket{\psi}$, while the other vectors are orthogonal to it. 

Fig.\ \ref{figSM1} shows an application of the procedure discussed for both the parent state $\ket{\psi}=\ket{\Lambda_r}$, the rotated IM scar parent state (see main text), and $\ket{\psi}=\ket{\mathrm{vac}}$, the Rydberg vacuum state. Here we plot the overlaps $|\bra{E_F}\psi \rangle|^2$ with Floquet eigenstates $\{\ket{E_F}\}$ before and after the rotation procedure for $\lambda=30$ and $\omega_D^f=15.09954$ for $L=30$ in the sector with $K=0, \mathcal{P}=+1$. While Fig. \ref{figSM1}(a) and (b) show a clear contrast in the zero-mode subspaces for the quantity $|\bra{E_F}\psi \rangle|^2$, Fig. \ref{figSM1}(c) and (d) for the vacuum as a parent state seem visually identical. However, several zero modes overlap with the vacuum before the procedure (though only slightly, ~$10^{-6}$ or so). After the unitary rotation within $\mathcal{V}$, all these overlaps are absorbed within the dressed vacuum scar $\ket{Z_{\psi}}$. Importantly, the resulting state is independent of the particular orthonormal basis chosen within the degenerate zero-mode subspace and is uniquely determined, up to an overall phase, by the projection of the parent state.

\section{System size dependence of $W_0(\psi)$}
\label{sec_SM_G}

\begin{figure} \label{figSM6}
 \includegraphics[width=\linewidth]{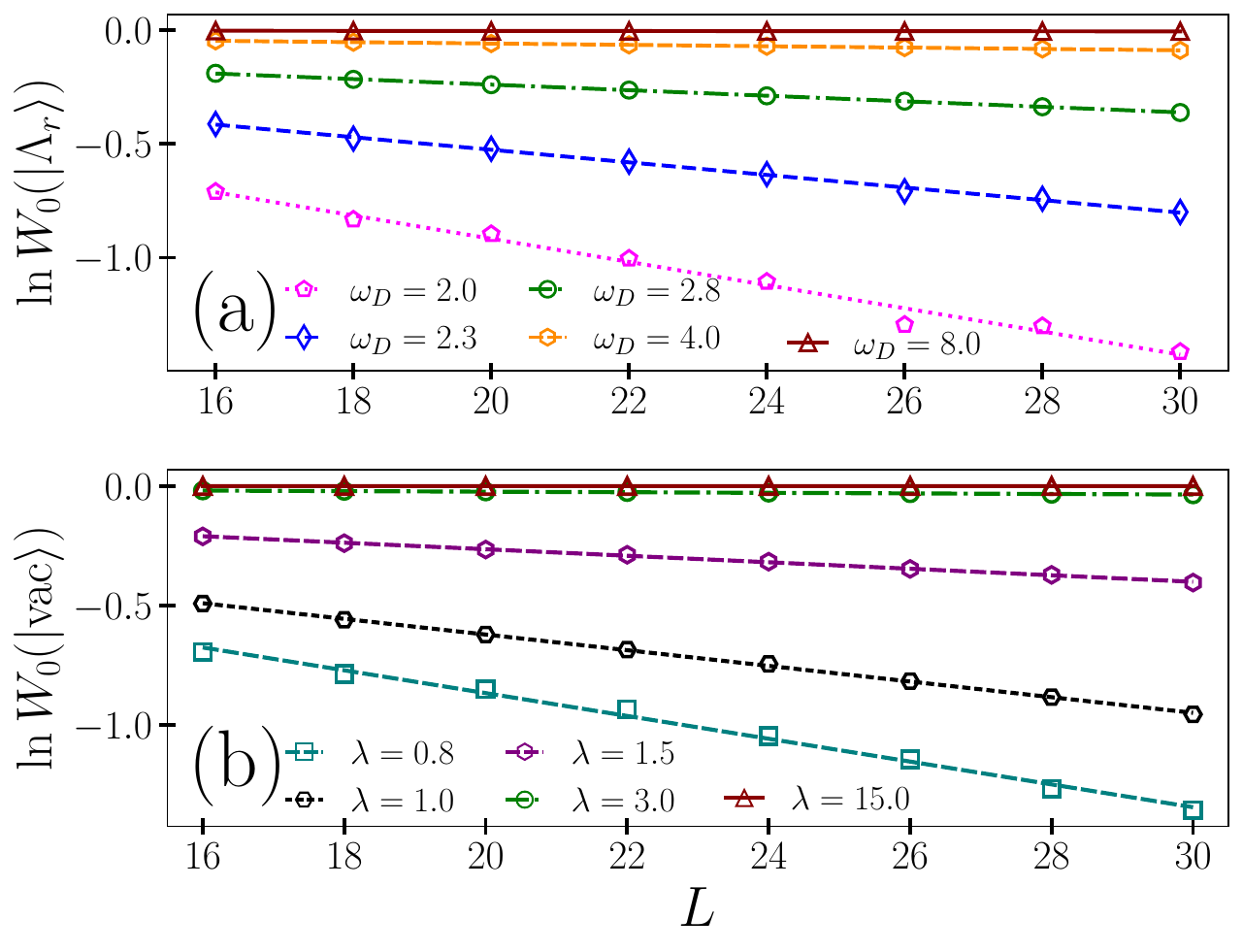}
 \caption{
 $\ln W_0(\ket{\psi})$ as a function of $L$ using $16 \leq L \leq 30$ for the parents: (a) $\ket{\psi}= \ket{\Lambda_r}$ at $x=1.2$ and several $\omega_D$ near the knee regime (see main text) and (b) $\ket{\psi}= \ket{\rm vac}$ for different $\lambda$ where $\omega_D=\omega^{\mathrm{f}}_D (\lambda)$ (see main text).  
 }
\end{figure}

To characterize the system-size dependence of the nullspace weight, we examine how $W_0(\ket{\psi})$ scales with $L$ for the two parent states considered in the main text. We characterize the absolute decay of the projection weight by fitting
\begin{equation}
W_0(\ket{\psi}) = A_0 \ e^{-\gamma_\psi L},
\label{eq:SM_W0_fit}
\end{equation}
where $\gamma_\psi$ denotes the decay exponent of the projection weight for the particular parent state (as discussed in the main text).

Fig.~\ref{figSM6} shows $\ln W_0(\ket{\psi})$ as a function of $L$ for the parent states, $\ket{\Lambda_r}$ and $\ket{\mathrm{vac}}$, respectively. For $\ket{\Lambda_r}$, we fix $x=1.2$ and consider
several drive frequencies in the vicinity of the knee regime (see main text). The data are approximately linear over the accessible system sizes $16\leq L\leq 30$, indicating an approximately exponential decay of the absolute projection weight. The extracted decay exponent increases continuously as $\omega_D$ is lowered from $\omega_D=8.0$ to $\omega_D=2.0$, with $\gamma_\psi$ increasing from $0.000194$ to $0.050958$ (Table~\ref{tab:W0_fitting_parameters}). Thus, the decrease in $W_0$ observed in the main text upon entering the knee regime is accompanied by an increasing finite-size decay rate.

For $\ket{\mathrm{vac}}$, we evaluate $W_0(\ket{\mathrm{vac}})$ at the drive frequency $\omega_D=\omega_D^{\mathrm f}(\lambda)$ associated with the maximum of $W_0$ for each $\lambda$. The corresponding system-size dependence is shown in Fig.~\ref{figSM6}(b). Again, the data are well described by an approximately exponential dependence over $16\leq L\leq 30$. The fitted decay exponent is very small at large drive amplitudes and increases as $\lambda$ decreases. In particular, $\gamma_\psi$ increases from approximately zero at $\lambda=15$ to $0.047836$ at $\lambda=0.8$. This behavior is consistent with the increasing dressing of the vacuum descendant as the system moves further from the regime where high-amplitude Floquet perturbation theory provides a controlled description.

The fitted values of $A_0$ and $\gamma_\psi$ are listed in Table~\ref{tab:W0_fitting_parameters}. The extracted values of $\gamma_\psi$ should not be treated as asymptotic values at $L \gg 1$, especially when $(\gamma_\psi)^{-1} \gg 30$, since the fits are performed over the finite range $16\leq L\leq30$. 

It is important to distinguish the decay exponent $\gamma_\psi$ in Eq.~\eqref{eq:SM_W0_fit} from the enhancement exponent $\gamma_e$ introduced in Eq.~(7) of the main text. The latter characterizes the projection weight relative to a typical state in the same symmetry sector,
\begin{equation}
\frac{W_0(\ket{\psi})}{W_0^{\mathrm{typ}}}
\sim e^{\gamma_e L},
\qquad
\gamma_e=\gamma_{\mathrm{typ}}-\gamma_\psi .
\label{eq:SM_gamma_e}
\end{equation}
Consequently, exponential decay of the absolute overlap $W_0(\ket{\psi})$ does not, by itself, imply the disappearance of the anomalous descendant. What is relevant is whether $\gamma_\psi$ is smaller than the corresponding typical decay exponent $\gamma_{\mathrm{typ}}$. The system-size dependence of $\ln[W_0/W_0^{\mathrm{typ}}]$, presented in the main text, directly probes this relative scaling and increases with $L$ over the accessible sizes in the parameter regimes discussed there.


\begin{table}[t]
\centering
\begin{tabular}{|c| c |c| c|}
\hline
Parent $\ket{\psi}$& Parameter & $A_0$ & $\gamma_\psi$ \\
\hline
\multirow{8}{*}{(a) $\ket{\Lambda_r}$}
& $\omega_D=8.0$ & $1.000006$ & $0.000194$ \\
& $\omega_D=5.0$ & $1.000083$ & $0.001238$ \\
& $\omega_D=4.0$ & $1.000355$ & $0.002975$ \\
& $\omega_D=3.0$ & $1.002373$ & $0.009250$ \\
& $\omega_D=2.8$ & $1.003977$ & $0.012189$ \\
& $\omega_D=2.6$ & $1.007356$ & $0.016453$ \\
& $\omega_D=2.3$ & $1.027139$ & $0.027635$ \\
& $\omega_D=2.0$ & $1.108011$ & $0.050958$ \\
\hline
\multirow{7}{*}{(b) $\ket{\mathrm{vac}}$}
& $\lambda=15.0$ & $1.000000$ & $0.000000$ \\
& $\lambda=5.0$ & $1.000041$ & $0.000079$ \\
& $\lambda=3.0$ & $1.000547$ & $0.001196$ \\
& $\lambda=2.0$ & $1.004312$ & $0.005925$ \\
& $\lambda=1.5$ & $1.008255$ & $0.013645$ \\
& $\lambda=1.0$ & $1.034965$ & $0.032829$ \\
& $\lambda=0.8$ & $1.093009$ & $0.047836$ \\
\hline
\end{tabular}
\caption{
Fitting parameters obtained from $\ln W_0(\ket{\psi})$ as a function of $L$ using $16 \leq L \leq 30$, assuming $W_0=A_0 e^{-\gamma_\psi L}$, for (a) $\ket{\psi}=\ket{\Lambda_r}$ at $x=1.2$ and several $\omega_D$ near the knee regime (see main text), and (b) $\ket{\psi}=\ket{\rm vac}$ for different $\lambda$, where $\omega_D=\omega_D^{\mathrm{f}}(\lambda)$ (see main text).
}
\label{tab:W0_fitting_parameters}
\end{table}


\section{Floquet perturbation theory for the driven PXP model}
\label{sec_SM_B}
We begin with the following time-dependent Hamiltonian given by 
\begin{equation}
    H(t) = H_0(t) + V, \label{eqA1}
\end{equation}
where $H_0(t)$ is diagonal in the computational basis and has a period of $T$ ($H(t+nT) = H(t)$ for $n \in \mathbb{Z}$).  Here $V$ has off-diagonal matrix elements in the computational basis (the PXP term in our case); it can be thought of as a perturbation when the magnitude of $V$ is small compared to that of $H_0$. In the absence of perturbation, the time evolution of a quantum state (from an initial time `$t_0$' to a later time `$t$') is governed by $H_0$, {\it i.e.},
\begin{align}
    \ket{\psi_0(t)} = U_0(t,t_0) \ket{\psi_0(t_0)} \label{eqA2}
\end{align}
which is the solution of Schr\"{o}dinger equation
\begin{align}
    i  \frac{\partial}{\partial t} \ket{\psi_0(t)} = H_0(t) \ket{\psi_0(t)} \   \label{eqA3}  
\end{align}
Here we work in units where $\hbar=1$.

Next, we define a state in interaction picture $\ket{\psi_I(t)} := U_0^\dagger(t,t_0) \ket{\psi(t)}$ , which satisfies the differential equation
\begin{align}
    i \, \frac{\partial}{\partial t} \ket{\psi_I(t)} = V_I(t) \ket{\psi_I(t)}
    \label{eqA4}
\end{align}
with $V_I(t)= U_0^\dagger(t,t_0) \, V  \, U_0(t,t_0)$. The corresponding unitary $U_I(t,t_0)$ can be obtained by solving the differential equation
\begin{align}
    i   \frac{\partial}{\partial t} U_I(t,t_0) = V_I(t) \, U_I(t,t_0)
    \label{eqA5}
\end{align}
Using the initial condition $U_I(t_0,t_0) = \mathbbm{1}$, one obtains the Dyson series
\begin{equation} \label{eqA6}
\begin{aligned}
 U_I(t,t_0) = & \mathbbm{1} + \left(- \, i  \right) \int_{t_0}^t dt' \, V^I(t')  \\ & \  + \ {\left(- \, i  \right)}^2 \int_{t_0}^{t} dt_1 \, V^I(t_1) \int_{t_0}^{t_1} dt_2 \, V^I(t_2) \ + \, \cdots
\end{aligned}
\end{equation}
The full time evolution is then given by the operator $U(t,t_0) = U_0(t,t_0) \ U_I(t,t_0)$. For our driven problem
\begin{align}
H_0 (t) = \frac{\lambda(t)}{2} \, \sum_j \sigma_j ^z \ \ , \quad \quad V = -w_0 \, \sum_j \tilde{\sigma}_j^x,  \label{eqA7}
\end{align}
where the piecewise function $\lambda(t)$ has the profile
\begin{equation}
\label{eqA8}
\lambda(t)=
\begin{cases}
-\lambda, & 0 \le t < T/2, \\
\lambda,  & T/2 \le t < T .
\end{cases}
\end{equation}
In the absence of a time-independent perturbation $V$, the `free' propagator is given by 
\begin{eqnarray}
U_0(t,0) &=& \exp{\left(-i \, \int_0^t dt' \, H_0(t')\right)} \nonumber \\
&=& \prod_j \, \exp{\left(-i \, \frac{\theta(t)}{2} \, \sigma_j^z\right)} \label{eqA9}
\end{eqnarray}
where we have introduced  $ \theta(t) := \, \int_0^t dt' \, \lambda(t')$ , where 
\begin{align}
\theta(t) =
\begin{cases}
- \lambda t \ , & 0 \leq t < T/2,  \label{eA10} \\
 \lambda (t-T) \ , & T/2 \leq t < T.
\end{cases}
\end{align}
In the interaction picture, the $V$ term takes the form 
\begin{align*}
V_I(t) &= U_0^\dagger(t,0) \, V \, U_0(t,0)\\
&= -w_0 \, \sum_j \ \left( e^{i \, \frac{\theta(t)}{2} \, \sigma_j^z} \right) \ \tilde{\sigma}_j^x \ \left(e^{-i \, \frac{\theta(t)}{2} \, \sigma_j^z} \right) \ .
\end{align*}
Using the following identity 
\begin{align*} 
e^{i \, \frac{\phi}{2} \, \sigma^z} \, \sigma^x  \, e^{-i \, \frac{\phi}{2} \, \sigma^z} = \cos{\phi} \ \sigma^x \ - \ \sin{\phi} \ \sigma^y  \ ,
\end{align*}
then leads to 
\begin{align}
V_I(t) = -w_0 \, \sum_j \left( \cos{\theta(t)}\, \tilde{\sigma}_j^x \ - \ \sin{\theta(t)}\, \tilde{\sigma}_j^y \right) \label{eqA11}
\end{align}
Next, we compute terms in the Dyson series and, from there, the effective Floquet Hamiltonian $H_F$ at each order in $w_0$.
Writing $$ U(T,0):= e^{-i T H_F}= \mathbbm{1} + \, \sum_{n=1}^{\infty} \alpha^n U_n \ ,$$
gives the effective Floquet Hamiltonian 
\begin{align*}
H_F = \sum_{n=0}^{\infty} \alpha^n H_F^{(n)} = \frac{i}{T} \, \ln{(\mathbbm{1} + \, \sum_{n=1}^{\infty} \alpha^n U_n)} \ .
\end{align*}
We then equate powers of $\alpha$ from both sides to obtain the following results.
\begin{equation} \label{eqA12}
\begin{aligned}
H_F^{(0)} &= 0, \\
H_F^{(1)} &= \frac{i}{T} \, U_1, \\
H_F^{(2)} &= \frac{i}{T} \Big( U_2 - \frac{1}{2} U_1^2 \Big), \\
H_F^{(3)} &= \frac{i}{T} \Big( U_3 - \frac{1}{2} ( U_1 U_2 + U_2 U_1 ) + \frac{1}{3} U_1^3 \Big), \\
& \ \vdots
\end{aligned}
\end{equation}
with the Dyson terms
\[
\begin{aligned}
U_1 &= (-i) \int_0^T dt_1 \, V_I(t_1), \\
U_2 &= (-i)^2 \int_0^T dt_1 \int_0^{t_1} dt_2 \, V_I(t_1) V_I(t_2), \\
U_3 &= (-i)^3 \int_0^T dt_1 \int_0^{t_1} dt_2 \int_0^{t_2} dt_3 \, V_I(t_1) V_I(t_2) V_I(t_3), \\ 
& \ \vdots
\end{aligned}
\]
Writing $U(T,0) = \exp(\Omega)$, we see that $\Omega_1 = U_1$ so that $H_F^{(1)}$ is given by the time-averaged Hamiltonian
\begin{align}
H_F^{(1)} = \frac{i}{T} \, \Omega_1 = \frac{1}{T} \, \int_0^T dt \, V_I(t) \ .  \label{eqA13}
\end{align}

Calculating $\Omega_2$, we obtain 
\begin{align*}
& \quad \Omega_2 = U_2 - \frac{1}{2} \, U_1^2 \\
&= (-i)^2 \, \left[ \int_{\Delta} \, V_1 V_2 \ - \ \frac{1}{2} \int_0^T dt_1 \, \int_0^{T} dt_2 \ V_I(t_1) \, V_I(t_2)\right] \\
&= - \left[ \int_{\Delta} V_1 V_2 \ - \ \frac{1}{2} \left( \int_\Delta V_1 V_2 \ + \ \int_\Delta V_2 V_1 \right)\right] \\
&= - \frac{1}{2} \, \int_\Delta [V_1, V_2] \ .
\end{align*}
where we denote `$\int_{\Delta} $' as a time-ordered integral, i.e.
\begin{align*}
\int_{\Delta} \, V_i V_j := \int_0^T dt_i \, \int_0^{t_i} dt_j \, V_I(t_i) \, V_I (t_j) \ .
\end{align*}
We thus obtain
\begin{align}
H_F^{(2)} = \frac{i}{T} \, \Omega_2 = - \frac{i}{2T} \, \int_0^T dt_1 \, \int_0^{t_1} dt_2 \, [V_I(t_1), V_I(t_2)] \ .  \label{eqA14}
\end{align}

For $\Omega_3$, we can similarly show
\begin{align*}
\Omega_3 &= U_3 \ - \ \frac{1}{2}\, \left( U_1U_2 + U_2 U_1 \right) \ + \ \frac{1}{3} \, U_1^3 \\
&= \frac{i}{6} \, \int_\Delta \left(\,  [V_1, [V_2,V_3]] \ + \ [[V_1, V_2], V_3]\, \right)
\end{align*}
and hence
\begin{align}
H_F^{(3)} &= \frac{i}{T}\, \Omega_3  =  -\frac{1}{6T} \, \int_0^T dt_1 \, \int_0^{t_1} dt_2 \, \int_0^{t_2} dt_3 \,  \nonumber \\& \Big( \, \left[ V_I(t_1), [V_I(t_2),V_I(t_3)]\right] \nonumber \\ & \qquad + \  \left[[V_I(t_1), V_I(t_2)], V_I(t_3) \right]\,\Big) \ .  \label{eqA15}
\end{align}
We club the results together in the following:
\begin{equation} \label{eqA16}
\begin{aligned}
H_F^{(1)} &= \frac{1}{T} \int_0^T dt_1 \, V_I(t_1), \\[2mm]
H_F^{(2)} &= -\frac{i}{2T} \int_0^T dt_1 \int_0^{t_1} dt_2 \, [V_I(t_1), V_I(t_2)], \\[2mm]
H_F^{(3)} &= \frac{i}{T}\, \Omega_3  =  -\frac{1}{6T} \, \int_0^T dt_1 \, \int_0^{t_1} dt_2 \, \int_0^{t_2} dt_3 \,  \\& \Big( \, \left[ V_I(t_1), [V_I(t_2),V_I(t_3)]\right]  \  + \  \left[[V_I(t_1), V_I(t_2)], V_I(t_3) \right]\,\Big) 
\end{aligned}
\end{equation}

Next, we note that 
\begin{align*}
\int_0^T dt \,  V_I(t)
&:= -w_0 \, \sum_j \left( I_c \,\tilde{\sigma}_j^x \ - \ I_s \, \tilde{\sigma}_j^y \right),\\
I_c &:= \int_0^T dt \, \cos{\theta(t)} 
= \frac{4}{\lambda} \, \sin\!\left(\frac{\lambda T}{4}\right)
      \cos\!\left(\frac{\lambda T}{4}\right) \ ,  \\
I_s &:= \int_0^T dt \, \sin{\theta(t)} 
= -\frac{4}{\lambda} \, \sin^2\!\left(\frac{\lambda T}{4}\right) \ .
\end{align*}
These identities allow us to obtain
\begin{align*}
H_F^{(1)} &= \, -\frac{4 w_0}{\lambda T} \, \sin{\left( \frac{\lambda T}{4}\right)} \\
&   \qquad \times \  \sum_j \left( \cos{\left( \frac{\lambda T}{4}\right)} \, \tilde{\sigma}_j^x  \ + \ \sin{\left( \frac{\lambda T}{4}\right)} \, \tilde{\sigma}_j^y  \right) \ .
\end{align*}
In terms of $\gamma := \lambda T/4 = x \pi$ (where $x$ is defined in the main text), we therefore find
\begin{align}
H_F^{(1)} = -w_0 \frac{\sin{\gamma}}{\gamma} \, \sum_j \left( \cos{\gamma} \ \tilde{\sigma}_j^x \ + \ \sin{\gamma} \ \tilde{\sigma}_j^y \right) \label{eqA17}
\end{align}
so that $\gamma/\pi= m\in \mathbb{Z^{+}}$ gives special points in parameter regime where $H_F^{(1)}$ vanishes.

Before we proceed further, we note down some relations which will be useful for computing higher-order terms in $H_F$. These are given by 
\begin{equation}\label{eqA18}
\begin{aligned}
P_k \, \sigma_k^x = \sigma_k^x \, P_k^\uparrow = \sigma_k^- \,,\quad &
\sigma_k^x \, P_k = P_k^\uparrow \, \sigma_k^x = \sigma_k^+ \,,\\[3pt]
P_k \, \sigma_k^y = \sigma_k^y \, P_k^\uparrow = i\,\sigma_k^- \,,\quad &
\sigma_k^y \, P_k = P_k^\uparrow \, \sigma_k^y = -\,i\,\sigma_k^+ \\
\bigl ( \sigma_k^+ \, \sigma_{k+1}^- \, + \, \sigma_k^- \, \sigma_{k+1}^+ \bigr) &= \frac{1}{2} \, \bigl(\sigma_k^x \,  \sigma_{k+1}^x \, + \, \sigma_k^y \, \sigma_{k+1}^y \bigr) \\
   \bigl ( \sigma_k^+ \, \sigma_{k+1}^- \, - \, \sigma_k^- \, \sigma_{k+1}^+ \bigr) &= \frac{i}{2} \, \bigl(\sigma_k^y \,  \sigma_{k+1}^x \, - \, \sigma_k^x \, \sigma_{k+1}^y \bigr)
\end{aligned}
\end{equation}\\
where $P_k = \sigma_k^- \ \sigma_k^+$ and $P_k^{\uparrow} = \sigma_k^+ \ \sigma_k^-$ are the down-spin and up-spin projectors respectively. We abbreviate $\theta(t_j)$ as $\theta_j$. Then using equation \eqref{eqA11}, we see
\begin{align*}
&[V_I(t_1) , V_I(t_2)] \\ & = \sum_{j \, , \, j'} \, w_0^2 \Big[ \left( \cos{\theta_1} \ \tilde{\sigma}_j^x \ - \ \sin{\theta_1} \  \tilde{\sigma}_j^y  \right) \, , \\& \qquad  \qquad \qquad  \left( \cos{\theta_2} \ \tilde{\sigma}_{j'}^x \ - \ \sin{\theta_2} \  \tilde{\sigma}_{j'}^y\right) \Big] \ .
\end{align*}
Here we see that non-vanishing commutators can only appear for $j' = j \, , \ , j \pm 1$ and using the identities \eqref{eqA18}, one can show
\begin{equation}  \label{eqA19}
\begin{aligned}
&[ V_I(t_1), V_I(t_2)]  = \, i \, w_0^2 \, \sin{(\theta_1 - \theta_2)} \\ 
&  \quad \times \ \sum_j\, \Big[\, P_{j-2} \, \left( \sigma_{j-1}^x \, \sigma_j^x \ + \ \sigma_{j-1}^y \, \sigma_j^y \right)\, P_{j+1} \\ & \qquad \qquad \  + \ 2 \, \tilde{\sigma}_j^z \, \Big] \ .
\end{aligned}
\end{equation}
But one can easily verify (using Mathematica) that
\begin{align*}
\int_0^T dt_1 \, \int_0^{t_1} dt_2 \, \sin{(\theta(t_1) - \theta(t_2))} = 0 \ . 
\end{align*}
This leads to 
\begin{align}
H_F^{(2)} &=  -\frac{i}{2T} \int_0^T dt_1 \int_0^{t_1} dt_2 \, [V_I(t_1), V_I(t_2)] = 0  \ .  \label{eqA20}
\end{align}
This result may also be inferred from the existence of a chiral operator $\mathcal{C} := \Pi_{j=1}^{L} \, \sigma_j^z$ that anticommutes with $H_F$ and, as a consequence, $H_F$ does not contain any terms that are constructed of an even number of $\tilde{\sigma}^\pm$. From this it follows that $H_F^{(2n)} = 0$ identically $\forall \ \, n \in \mathbbm{Z^+}$. 

The higher-order nested commutator yields 
\begin{widetext} 
\begin{eqnarray}
[V_I(t_1), [V_I(t_2),V_I(t_3)]] &=& -[[V_I(t_2),V_I(t_3)], V_I(t_1)] = i \, w_0^3 \, \sin{(\theta_2 - \theta_3)} 
\Big( \, \sum_{j , \, j'} \, \Big[ \, P_{j-2} \, \left( \sigma_{j-1}^x \, \sigma_j^x \ + \ \sigma_{j-1}^y \, \sigma_j^y \right)\, \nonumber\\
&& \times P_{j+1} \left( \cos{\theta_1} \ \tilde{\sigma}_{j'}^x \ - \ \sin{\theta_1} \  \tilde{\sigma}_{j'}^y \right) \, \Big]
+ \ 2 \, \sum_{j , \, j'} \, \Big[ \, \tilde{\sigma}_{j'}^z \, , \, \left( \, \cos{\theta_1} \ \tilde{\sigma}_{j}^x \ - \ \sin{\theta_1} \  \tilde{\sigma}_{j}^y \, \right) \, \Big] \, \Big) \ .
\end{eqnarray}
\end{widetext} 
Noting the non-zero results for $j'=j-2\, , \, j\pm1 \, , \, j$ ; one gets 
\begin{widetext} 
\begin{eqnarray}
[V_I(t_1),[V_I(t_2),V_I(t_3)]] &=& -\,4\, w_0^3 \sin(\theta_2-\theta_3) \left\{ 
- \sum_j\Bigl( i \, e^{i\theta_1} \, \tilde{\sigma}_{j}^+ \ + \ \text{H.c.}\Bigr) + \sum_j P_{j-2} \, \Bigl( \,  ie^{i \theta_1} \, \sigma_{j-1}^+ \, \sigma_{j}^- \, \sigma_{j+1}^+  \ + \ \text{h.c.} \Bigr)\, \right. \nonumber\\
&& \times \left. P_{j+2}  \sum_j \, P_{j-1} \, \left( \,  i \, e^{i\theta_1} \, \tilde{\sigma}_{j+1}^+ \ + \ \text{h.c.}\, \right) \, \sum_j \, \left( \,  i \, e^{i\theta_1} \, \tilde{\sigma}_{j}^+ \ + \ \text{h.c.} \, \right) \, P_{j+2} \ \right\}. \label{eqA21}
\end{eqnarray}
\end{widetext} 
The other commutator $[[V_I(t_1),V_I(t_2)],V_I(t_3)]$  can be obtained on same token by replacing \ $\theta_{1,2,3} \to \theta_{3,2,1}$ in \eqref{eqA21}.

Referring to \eqref{eqA15}, we find the amplitude as follows.
\begin{align*}
&A_0 = -\frac{4\, w_0^3}{6T} \, \times \, \int_0^T dt_1 \int_0^{t_1} dt_2 \int_0^{t_2} dt_3 \\ 
& \quad  \ i \, \Big(\, \sin(\theta_1-\theta_2) \, e^{i \theta_3} \ - \  \sin(\theta_2-\theta_3) \, e^{i \theta_1} \, \Big) \ . 
\end{align*}
Using Mathematica, we carry out the integral to get
\begin{equation}
\begin{aligned} \label{eqA22}
A_0  &= \, \frac{w_0^3 e^{-i\lambda T}}{6 i \lambda^3 T} \, \times \\
&\quad  \left[ \, e^{3i\lambda T/2} +3 \, e^{i\lambda T/2} \, \left(1+ i \lambda T \right) +2 \, \left(1-3 e^{i\lambda T} \right) \, \right] \ . 
\end{aligned}
\end{equation}
We thus get the $3$rd-order Floquet Hamiltonian $H_F^{(3)}$
\begin{equation}
\begin{aligned}
H_F^{(3)} &=  \sum_j  P_{j-2} \left( A_0 \ \sigma_{j-1}^+ \, \sigma_j^- \, \sigma_{j+1}^+ \ + \ \text{h.c.} \right) P_{j+2} \\ & \ -   \sum_j \left( A_0 \, \tilde{\sigma}_j^+ \ + \ \text{h.c.} \right)  - \sum_j \left( A_0 \, \tilde{\sigma}_j^+ \ + \ \text{h.c.} \right)P_{j+2}  \\ & \ -  \sum_j P_{j-1} \, \left( A_0 \, \tilde{\sigma}_{j+1}^+ \ + \ \text{h.c.} \right) \ .  \label{eqA23}
\end{aligned}
\end{equation}
Calculating higher-order terms in FPT becomes increasingly cumbersome, and we do not compute $H_F^{(n)}$ for $n>3$ in this work. 

\section{Additional features of the dressed IM scar}
\label{sec_SM_D}

The dressed IM scar for the Floquet problem is formed by projecting the rotated IM scar parent state $\ket{\Lambda_r}=\mathcal{U}(x)\ket{\Lambda}$, where $\ket{\Lambda}$ is a highly entangled zero mode of the undriven PXP model (see main text), onto the nullspace of $H_F$ and performing the procedure discussed in Sec. \ref{sec_SM_A} to find the zero mode parallel to this projection. The main text already discussed the presence of anomalous two-point correlations between the antipodal spins in $|Z_{\Lambda_r} \rangle$ as long as $W_0 (\ket{\Lambda_r})$ is high compared to $W_0^{\mathrm{typ}}$, while the same correlator approaches its infinite-temperature value as $W_0 (\ket{\Lambda_r}) \simeq W_0^{\mathrm{typ}}$. Here, we present further evidence for the same.

\begin{figure*}
    \includegraphics[width=\linewidth]{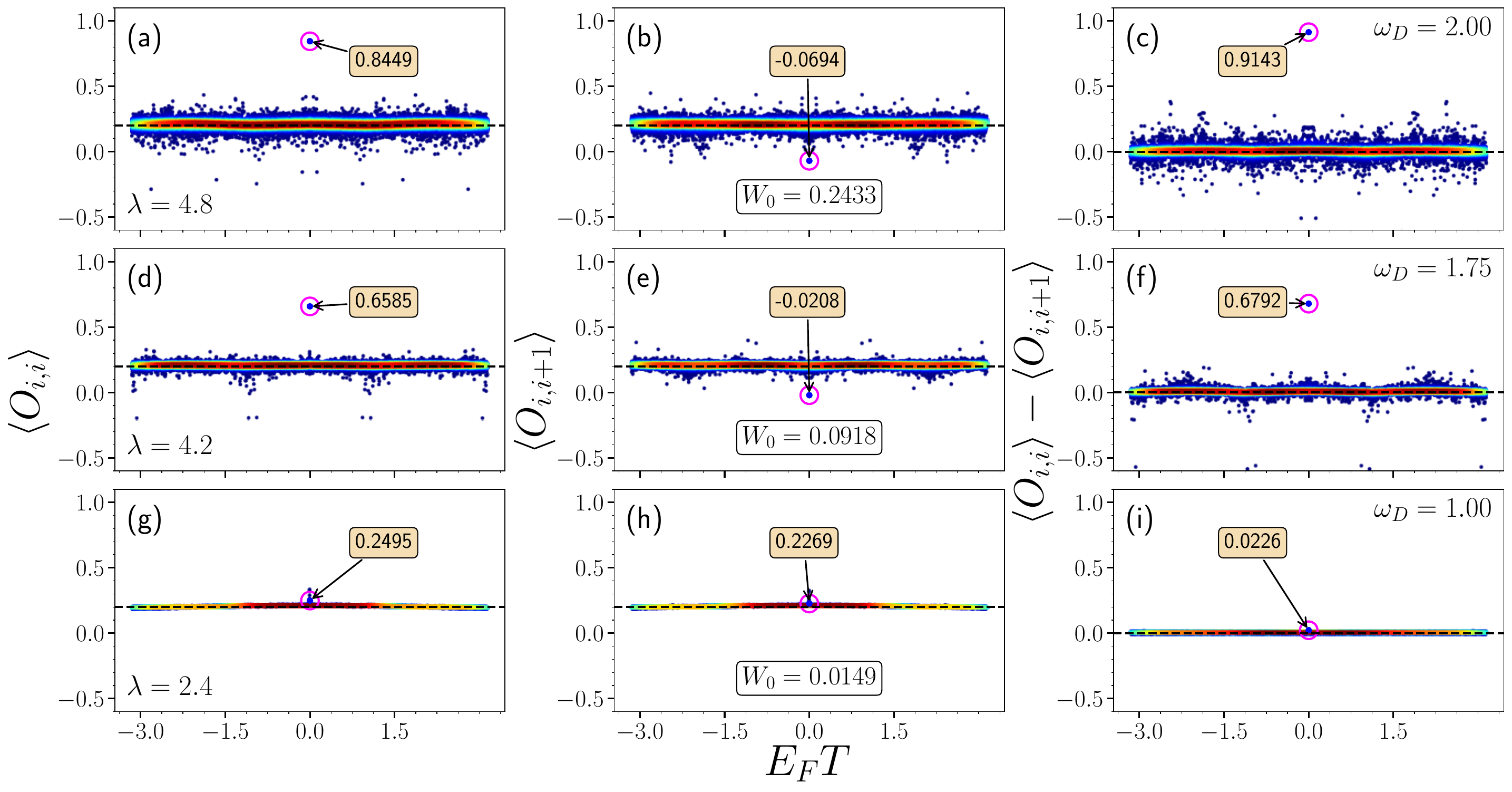}
    \caption{$\langle \mathcal{O}_{i,i}\rangle$ (first column), $\langle \mathcal{O}_{i,i+1}\rangle$ (middle column) and $\langle \mathcal{O}_{i,i}\rangle - \langle \mathcal{O}_{i,i+1}\rangle$ (last column) with respect to the Floquet eigenstates $\{\ket{E_F}\}$ for the system size $L=30$ in $K=0,\mathcal{P}=+1$ sector for $(\lambda, \omega_D)$ values $(4.8, 2.0)$ (first row (a),(b),(c)), $(4.2, 1.75)$ (middle row (d),(e),(f)) and $(2.4, 1.0)$ (last row (g),(h),(i)) keeping $x=1.2$ to be fixed. The dressed IM scar is circled in magenta in all panels, and the corresponding correlator value is highlighted and written within each subplot for this state. The infinite-temperature ensemble average for $\langle \mathcal{O}_{i,j}\rangle$ for two sites that are well-separated equals $1/5$ for $L \gg 1$ and is shown by a black dashed horizontal line for the first two columns, whereas in the last column, the difference is shown by the same at $0$. The density of states uses the same color map in all three panels, where warmer colors correspond to higher density of states. 
}
    \label{figSM2}
\end{figure*}

Fig. \ref{figSM2} displays $\langle\mathcal{O}_{i,j}\rangle$ for $j=i$ (left-most panels), for $j=i+1$ (middle panels), and the difference of these two operators (right-most panels) for each of the Floquet eigenstates in the $K=0, \mathcal{P}=+1$ sector for a fixed $x=1.2$ using $L=30$. Here the drive frequency $\omega_D$ is varied from $\omega_D=2.0$ (upper row), to $\omega_D=1.75$ (middle row) to $\omega_D=1.0$ (bottom row). While the values of $\langle\mathcal{O}_{i,j}\rangle$ for $j=i$ (left-most panels) and for $j=i+1$ (middle panels) deviate from $1/5$ for a dressed IM scar as discussed in the main text, the difference between these two correlators (right-most panels) provides an additional diagnostic of the nonthermal spatial structure inherited from the IM parent state, with an expected infinite-temperature value approaching zero. We circle the dressed IM scar $\ket{Z_{\Lambda_r}}$ in magenta in all panels of Fig.~\ref{figSM2} for clarity. From Fig. 2 of the main text, it is clear that while $\omega_D=2.0$ and $\omega_D=1.75$ correspond to the knee regime where $W_0 \gg W_0^{{\mathrm{typ}}}$; in contrast, $\omega_D=1.0$ corresponds to a drive frequency in the second plateau where $W_0 \approx W_0^{\mathrm{typ}}$ at a fixed $x=1.2$. 

The calculated $W_0$ for the three drive frequencies are also shown in the middle column of Fig.~\ref{figSM2} and correspond to $W_0 =0.2433$ for $\omega_D=2.0$, $W_0=0.0918$ for $\omega_D=1.75$, and $W_0=0.0149$ for $\omega_D=1.0$ for $L=30$. We note here that $W_0^{\mathrm{typ}} =0.0132$ for $L=30$ from the formula derived in the main text. While a large majority of the Floquet eigenstates have $\langle\mathcal{O}_{i,j}\rangle \approx 1/5$ for $j=i$ and $j=i+1$, and $\langle\mathcal{O}_{i,i}- \mathcal{O}_{i,i+1}\rangle \approx 0$ for all the three drive frequencies displayed in Fig.~\ref{figSM2}, the correlators of $|Z_\psi \rangle$ for $|\psi \rangle=\ket{\Lambda_r}$ are visible as clear outliers for $\omega_D=2.0$ (Fig.~\ref{figSM2} (a), (b), (c)) and $\omega_D=1.75$ (Fig.~\ref{figSM2} ((d), (e), (f)); the corresponding values of the correlators deviate more from the expected values for $|\Lambda_r \rangle$ as $\omega_D$ is lowered. However, for $\omega_D =1.0$ (Fig.~\ref{figSM2} ((g), (h), (i)), the correlators of $|Z_\psi \rangle$ cannot be distinguished anymore from the corresponding values for the typical Floquet eigenstates within finite-size fluctuation effects and cease to be outliers. 

Another point about the dressed IM scar can be illustrated by using ED to numerically compute $R[H_F^{\mathrm{rot}}]$, where $R[A] =  \Vert {\rm Im}[A] \Vert_F / \Vert {\rm Re}[A] \Vert_F$, for $x=1.2$ and $L=28$ as a function of $\omega_D$ in the sector $K=0, \mathcal{P}=+1$ ($\Vert \cdot \Vert_F$ denotes the Frobenius norm of a matrix). Here, we perform a unitary rotation, $\mathcal{U}(x)$ (see main text) to obtain $H_F^{\mathrm{rot}}:= \mathcal{U}^{\dagger}(x) H_F \, \mathcal{U}(x)$ for a non-integer $x$; this procedure ensures that $H_F^{(1)}$ is real symmetric in the computational basis (see main text). 

\begin{figure} \label{fig_fro_norm}
    \includegraphics[width=\linewidth]{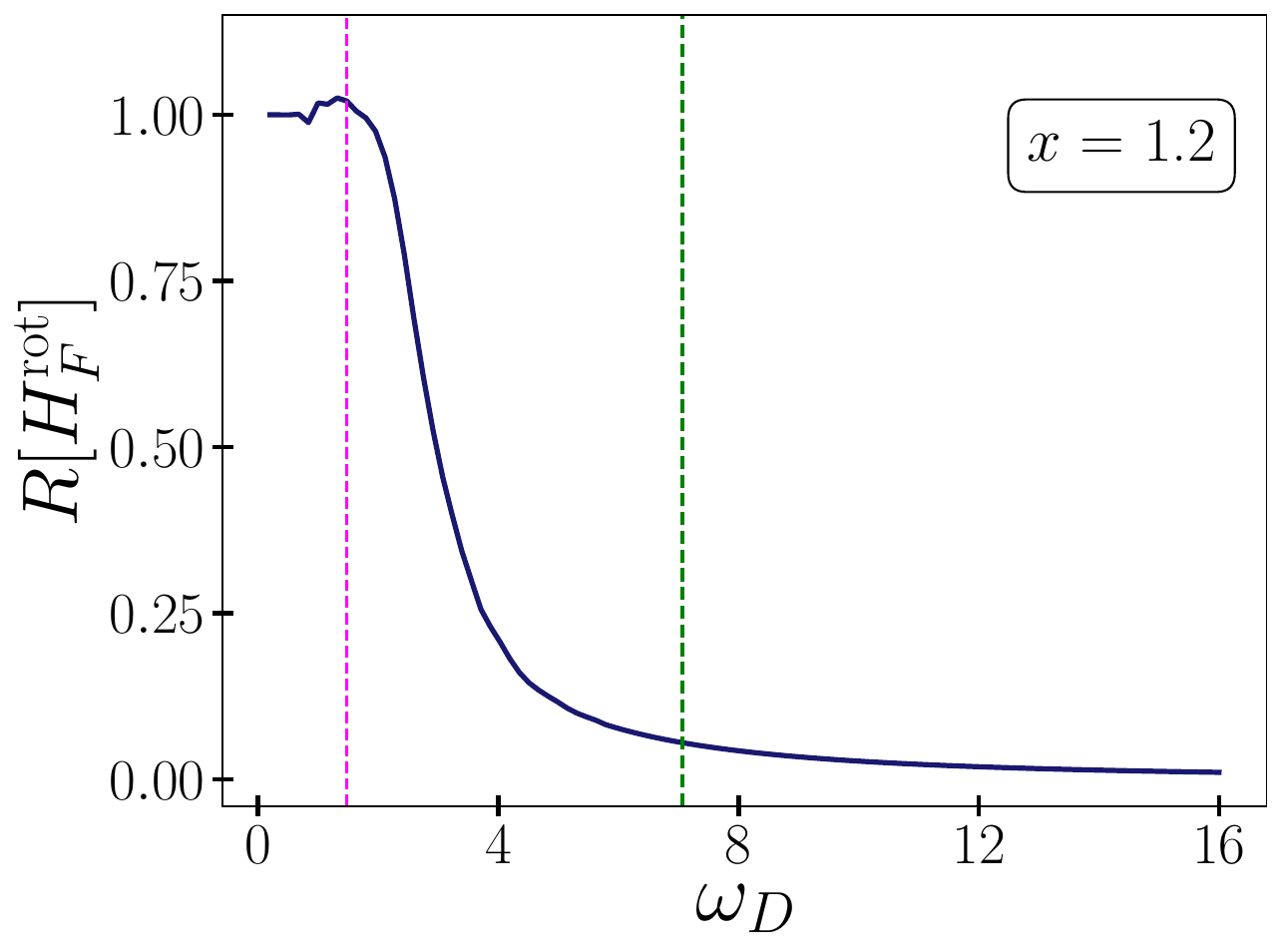}
    \caption{Plot of the $R[H^{\mathrm{rot}}_F]$ as a function of $\omega_D$ for $L=28$ in the $K=0,\mathcal{P}=+1$ sector using a fixed $x=1.2$. The vertical dotted lines (magenta and green) show the onset of the two plateaus at $\omega_D \approx 1.48$ and $\omega_D \approx 7.06$ (main text, end matter).}
\end{figure}

Fig.~\ref{fig_fro_norm} shows the corresponding results. Fig.~\ref {fig_fro_norm} indicates the onset of the two plateaus (main text, end matter) at $\omega_D \approx 7.0$ and $\omega_D \approx 1.5$ with dashed lines. We see that $R$ displays a monotonic rise from $R \approx 0$ at large $\omega_D$ until the onset of the second plateau at $\omega_D \approx 1.5$, where $W_0(\ket{\Lambda_r}) \simeq W_0^{\mathrm{typ}}$ and $|Z_{\Lambda_r} \rangle$ cannot be distinguished from a typical mode in the Floquet nullspace. Below $\omega_D \sim 1.5$, $R$ shows fluctuations around $1$. We expect $R \approx 1$ for a generic random complex Hermitian matrix. The behavior of $R$, thus, suggests that the anomalous correlations in the dressed IM scar survives as long as the real part of $H_F$ ``dominates" over the imaginary part of $H_F$ and melts away when they become equally important in a statistical sense as captured by this Frobenius norm based measure.  

\section{FPT terms that do not annihilate vacuum}
\label{sec_SM_E}

We note that in our model, $\{H_F, \mathcal{C} \}=0$; consequently, $H_F$ acting on $\ket{\mathrm{vac}}=\ket{\downarrow \cdots \downarrow \downarrow}$ can only generate a linear combination of Fock states where each of them has an odd number of up-spins.  Calculating $H_F$ perturbatively to third order using FPT already leads to an effective single-spin flip term starting from $\ket{\mathrm{vac}}$, which we represent by $w_{\mathrm{eff}}=w^{(1)}+w^{(3)}+\cdots$ (see main text), that generates Fock states with a single flipped spin such as  $\ket{\downarrow\cdots \downarrow \uparrow \downarrow \cdots}$ and its translated copies. This $w_{\mathrm{eff}}$ term gets contributions from all higher-order terms in FPT such as $H_F^{(5)}$, etc., which the ellipsis represents, but we have not calculated these analytically at large $\lambda$. 

Considering higher-order terms in $H_F$ also leads to non-zero connections between vacuum and Fock states with three up-spins (such that the up-spins have the smallest separation between them without violating the strong Rydberg blockade constraints) such as $\ket{\downarrow\cdots \downarrow \uparrow \downarrow \uparrow \downarrow \uparrow \downarrow \cdots}$ and its translated copies. Other Fock states where the three up-spins are farther apart are generated at even higher orders, simply because $H(t)$ is local in real space. Going to even higher orders in FPT leads to terms in $H_F$ that generate (acting on vacuum) Fock states with five close-by up-spins such as $\ket{\downarrow \cdots \downarrow \uparrow \downarrow \uparrow \downarrow \uparrow \downarrow \uparrow \downarrow \uparrow \downarrow \cdots }$ and its translated copies. Here, we will show that the terms that lead to $3$ ($5$) up-spin Fock states can only arise at or beyond $7^{\mathrm{th}}$ ($13^{\mathrm{th}}$) order in FPT using properties like locality and the constrained nature of the Hilbert space.  

We begin by writing $V_I(t) = \sum_j v^{\pm}_j(t)$ as a sum over local terms, each of which has support on $3$ consecutive sites (\ref{eqA11}), where $v^{\pm}_j$ can be interpreted as $P_{j-1}\sigma^{\pm}_jP_{j+1}=\tilde{\sigma}_j^{\pm}$, up to a factor of $(w_0/\gamma) \sin{\gamma} \, e^{\mp i\gamma}$. At $n$-th order, the different terms of the FPT can then be thought of as different ``operator strings" $\prod_{a=1}^n \tilde{\sigma}_{i_a}^{p_{i_a}} $; where the sites $i_1, \cdots,i_n$ need not be distinct from each other and $p_{j}=+$ or $-$ for each $i_1, \cdots, i_n$. However, the sites $i_1, \cdots, i_n$ should not be disconnected on the lattice, since such terms vanish due to the nested commutator structure of the FPT terms. 

We now ask for the maximum ``operator length", $\ell_{\rm max}$,  of such an operator string on the lattice at the $n$-th order. It is clear that $\ell(v_j^{\pm})=2=\ell_{{\rm max}}[v_j^{\pm}]$, where the unit of length is measured in terms of the lattice spacing. Now, consider $X = [v^{\pm}_{j'},v^{\pm}_j]$. It is easy to see that $j'$ can at best be $j \pm 1$ for a non-zero commutator since $j'=j \pm 2$ leads to zero, as these operators have down-spin projectors on both ends that commute with each other. Thus, it follows that ${\ell}_{\mathrm{max}}(X)=3$. Furthermore, these length-$3$ operators will have down-spin projectors on both ends by construction. A recursive extension of this argument gives 
\begin{equation}
    {\ell}_{\mathrm{max}}(H_F^{(n)})=n+1
    \label{eqLmax}
\end{equation}
where $\ell_{\mathrm{max}} (\mathcal{O})$ denotes the length of an operator $\mathcal{O}$ generated at $n$-th order in FPT.

Next, we look for operator strings that do not annihilate the vacuum state. These operator strings must contain $m$ distinct $\tilde{\sigma}_{i_1},\cdots,\tilde{\sigma}_{i_m}^+$ with distinct sites, with the restriction that $m$ is odd. As argued earlier, this restriction follows from the anti-commutation of $H_F$ with $\mathcal{C}$ (see main text). Furthermore, these sites cannot be nearest neighbors to satisfy the Hilbert space constraints. We call such operators ``lone" operators. Note that $\tilde{\sigma}^-$ cannot be a ``lone" operator since such an operator annihilates the vacuum state. The operator string can also contain ``pairs" of the form $\tilde{\sigma}_k^- \ \tilde{\sigma}_k^+$ or $\tilde{\sigma}_k^+ \ \tilde{\sigma}_k^-$, where the pairs can act on the remaining sites as well as the sites of the lone operators, where there is no restriction for these sites to be distinct. 

Let us now consider an operator string with $m$ lone operators and $q$ pair operators. At $n$-th order of FPT, which corresponds to $n$ insertion of $V_{I}$'s, we have various operator strings each consists of $n$ number of $\tilde{\sigma}^{\pm}$ in total, which implies that 
\begin{equation}
\label{eqCons}
  n=m+2q.  
\end{equation}
While operator strings can have pair operators on the same sites, making their operator lengths fluctuate, none of these operator strings can have an operator length bigger than $(n+1)$ (Eq.~\ref{eqLmax}) or smaller than $2$. Noting the following operator identities $\tilde{\sigma}_j^-\tilde{\sigma}_j^+ = P_{j-1}P_{j}P_{j+1}$ and $\tilde{\sigma}_j^+\tilde{\sigma}_j^- = P_{j-1}P^{\uparrow}_{j}P_{j+1}$ (where $P^{\uparrow}_j$ denotes an up-spin projector on site $j$), one can prove the following equalities at the operator level:
\begin{align*}
    \left(\tilde{\sigma}_j^+ \tilde{\sigma}_j^- \right) \tilde{\sigma}_j^+ &= \tilde{\sigma}_j^+ \left(\tilde{\sigma}_{j}^- \tilde{\sigma}_j^+ \right) = \tilde{\sigma}_j^+ \\
    \left(\tilde{\sigma}_j^- \tilde{\sigma}_j^+ \right) \tilde{\sigma}_j^+ &= \tilde{\sigma}_j^+ \left(\tilde{\sigma}_{j}^+ \tilde{\sigma}_j^- \right) = 0
\end{align*}
and also
\begin{align*}
\left(\tilde{\sigma}_j^+ \tilde{\sigma}_j^- \right) \tilde{\sigma}_{j\pm1}^+ &= 0 = \left(\tilde{\sigma}_j^- \tilde{\sigma}_j^+ \right) \tilde{\sigma}_{j\pm1}^+ \\
\tilde{\sigma}_{j\pm1}^+\left(\tilde{\sigma}_j^+ \tilde{\sigma}_j^- \right) &= 0 \\
\tilde{\sigma}_{j\pm1}^+\left(\tilde{\sigma}_j^- \tilde{\sigma}_j^+ \right) &= P_{j\pm2} \, \tilde{\sigma}_{j\pm1}^+
\end{align*}
This along with the fact that a lone operator at site $j$ and a pair operator at site $j'$ commute if $|j'-j|\geq 2$ implies that operator strings that do not annihilate the vacuum state can always be cast to a ``normal form"; this means that it can be expressed as a product of the pairs $\left(\tilde{\sigma}_j^- \tilde{\sigma}_j^+ \right)$ (and not the other variety of pairs), followed by a product of lone $\tilde{\sigma}$ at the extreme left, when acting on $\ket{\mathrm{vac}}$:
\begin{align}
\left(\prod_{a=1}^{m}\tilde{\sigma}_{i_a}^{+}\right)\left(\prod_{b=1}^{q} \left(\tilde{\sigma}_{j_b}^{-}\tilde{\sigma}_{j_b}^{+}\right)\right)\ket{\mathrm{vac}}.
\end{align}

For $n=1$, only possible solution of equation \eqref{eqCons} is $m=1\ , \ l=0$ and one gets $\tilde{\sigma}_j^+$. For $n=3$, the following properties are worth noting.
 \begin{align*}
     \sigma_k^+ \, P_k = \sigma_k^+  \ , \quad P_k \, \sigma_k^+ = 0 
 \end{align*}
Let us look at terms for $m=1 \ , \ q = 1$. \\
$(i)$
\begin{align*}
    & \tilde{\sigma}_j^+ \ \left( \, \tilde{\sigma}_{j+1}^- \ \tilde{\sigma}_{j+1}^+ \, \right) = \left(\, P_{j-1} \, \sigma_j^+ \, P_{j+1} \, \right) \ \left(\, P_{j} \, P_{j+1} \, P_{j+2} \, \right) \\
    &= P_{j-1} \, \sigma_j^+ \, P_{j+1} \, P_{j+2} = \tilde{\sigma}_j^+ \ P_{j+2}
\end{align*} \newline
$(ii)$
\begin{align*}
    & \tilde{\sigma}_j^+ \  \left( \, \tilde{\sigma}_{j-1}^- \ \tilde{\sigma}_{j-1}^+ \, \right)  = \left(\, P_{j-1} \, \sigma_j^+ \, P_{j+1} \, \right) \  \left(\, P_{j-2} \,   P_{j-1} \, P_{j} \, \right)   \\
    &= P_{j-2} \, P_{j-1} \, \sigma_j^+ \, P_{j+1} = P_{j-2} \ \tilde{\sigma}_j^+
\end{align*} \newline
$(iii)$ 
\begin{align*}
    & \tilde{\sigma}_j^+ \ \left( \, \tilde{\sigma}_{j}^- \ \tilde{\sigma}_{j}^+ \, \right) = \left(\, P_{j-1} \, \sigma_j^+ \, P_{j+1} \, \right) \ \left(\, P_{j-1} \, P_{j} \, P_{j+1} \, \right) \\
    &= P_{j-1} \, \sigma_j^+ \, P_{j+1}  = \tilde{\sigma}_j^+ 
\end{align*}
These are indeed the local terms obtained at 3rd order \eqref{eqA23} which do not annihilate the vacuum. Next, for $n=5$, we note the following terms coming from $m=1 \ , \ q = 2$. \\
$(i)$
\begin{align*}
    & \tilde{\sigma}_j^+ \  \left( \, \tilde{\sigma}_{j-1}^- \ \tilde{\sigma}_{j-1}^+ \, \right) \left( \, \tilde{\sigma}_{j+1}^- \  \tilde{\sigma}_{j+1}^+ \, \right) \\ &  = \left(\, P_{j-1} \, \sigma_j^+ \, P_{j+1} \, \right) \ \left(\, P_{j-2} \, P_{j-1} \, P_{j} \, \right)  \ \left( \, P_{j} \, P_{j+1} \, P_{j+2} \, \right) \\
    & = P_{j-2} \, P_{j-1} \, \sigma_{j}^+ \, P_{j+1} \, P_{j+2} = P_{j-2} \ \tilde{\sigma}_j^+ \ P_{j+2}
\end{align*}
\newline
$(ii)$
\begin{align*}
    & \tilde{\sigma}_j^+ \  \left( \, \tilde{\sigma}_{j}^- \ \tilde{\sigma}_{j}^+ \, \right) \left( \, \tilde{\sigma}_{j+1}^- \  \tilde{\sigma}_{j+1}^+ \, \right) \\ &  = \left(\, P_{j-1} \, \sigma_j^+ \, P_{j+1} \, \right) \ \left(\, P_{j-1} \, P_{j} \, P_{j+1} \, \right)  \ \left( \, P_{j} \, P_{j+1} \, P_{j+2} \, \right) \\
    & = P_{j-1} \, \sigma_{j}^+ \, P_{j+1} \, P_{j+2} = \, \tilde{\sigma}_j^+ \ P_{j+2}
\end{align*}
\newline
$(iii)$
\begin{align*}
    & \tilde{\sigma}_j^+ \  \left( \, \tilde{\sigma}_{j}^- \ \tilde{\sigma}_{j}^+ \, \right) \left( \, \tilde{\sigma}_{j-1}^- \  \tilde{\sigma}_{j-1}^+ \, \right) \\ &  = \left(\, P_{j-1} \, \sigma_j^+ \, P_{j+1} \, \right) \ \left(\, P_{j-1} \, P_{j} \, P_{j+1} \, \right)  \ \left( \, P_{j-2} \, P_{j-1} \, P_{j} \, \right) \\
    & = P_{j-2} \, P_{j-1} \, \sigma_{j}^+ \, P_{j+1}  = P_{j-2} \ \tilde{\sigma}_j^+ 
\end{align*}
\newline
$(iv)$
\begin{align*}
    & \tilde{\sigma}_j^+ \  \left( \, \tilde{\sigma}_{j}^- \ \tilde{\sigma}_{j}^+ \, \right) \left( \, \tilde{\sigma}_{j}^- \  \tilde{\sigma}_{j}^+ \, \right) \\ &  = \left(\, P_{j-1} \, \sigma_j^+ \, P_{j+1} \, \right) \ \left(\, P_{j-1} \, P_{j} \, P_{j+1} \, \right)  \ \left( \, P_{j-1} \, P_{j} \, P_{j+1} \, \right) \\
    & =  P_{j-1} \, \sigma_{j}^+ \, P_{j+1}  =  \ \tilde{\sigma}_j^+ 
\end{align*}
These are the only possible terms that give dynamics to the vacuum at $5$-th order, since $m=3 \ , \ q=1$ terms are identically zero due to `gaps' (lack of pairs) at intermediate sites, which leads to a disconnected operator string. Now for $n=7$, terms coming from $m=1 \ , \ q = 3$ are the following.\\
$(i)$
\begin{align*}
    & \tilde{\sigma}_j^+ \  \left( \, \tilde{\sigma}_{j-1}^- \ \tilde{\sigma}_{j-1}^+ \, \right) \left( \, \tilde{\sigma}_{j+1}^- \  \tilde{\sigma}_{j+1}^+ \, \right) \left( \, \tilde{\sigma}_{j+2}^- \  \tilde{\sigma}_{j+2}^+ \, \right)\  \\ &= \left(\, P_{j-1} \, \sigma_j^+ \, P_{j+1} \, \right) \ \left(\, P_{j-2} \, P_{j-1} \, P_{j} \, \right)  \ \left( \, P_{j} \, P_{j+1} \, P_{j+2} \, \right) \\ & \quad \left( \, P_{j+1} \, P_{j+2} \, P_{j+3} \, \right) \\
    & = P_{j-2} \, P_{j-1} \, \sigma_{j}^+ \, P_{j+1} \, P_{j+2} \, P_{j+3} = P_{j-2} \ \tilde{\sigma}_j^+ \ P_{j+2} \ P_{j+3}
\end{align*}
\newline
$(ii)$
\begin{align*}
    & \tilde{\sigma}_j^+ \  \left( \, \tilde{\sigma}_{j}^- \ \tilde{\sigma}_{j}^+ \, \right) \left( \, \tilde{\sigma}_{j+1}^- \  \tilde{\sigma}_{j+1}^+ \, \right) \left( \, \tilde{\sigma}_{j+2}^- \  \tilde{\sigma}_{j+2}^+ \, \right) \\ & = \left(\, P_{j-1} \, \sigma_j^+ \, P_{j+1} \, \right) \ \left(\, P_{j-1} \, P_{j} \, P_{j+1} \, \right)  \ \left( \, P_{j} \, P_{j+1} \, P_{j+2} \, \right) \\ & \quad \left( \, P_{j+1} \, P_{j+2} \, P_{j+3} \, \right)\\
    & = P_{j-1} \, \sigma_{j}^+ \, P_{j+1} \, P_{j+2} \, P_{j+3} = \, \tilde{\sigma}_j^+ \ P_{j+2} \ P_{j+3}
\end{align*}
\newline
$(iii)$
\begin{align*}
    & \tilde{\sigma}_j^+ \  \left( \, \tilde{\sigma}_{j-1}^- \ \tilde{\sigma}_{j-1}^+ \, \right) \left( \, \tilde{\sigma}_{j+1}^- \  \tilde{\sigma}_{j+1}^+ \, \right) \left( \, \tilde{\sigma}_{j-2}^- \  \tilde{\sigma}_{j-2}^+ \, \right)\  \\ &= \left(\, P_{j-1} \, \sigma_j^+ \, P_{j+1} \, \right) \ \left(\, P_{j-2} \, P_{j-1} \, P_{j} \, \right)  \ \left( \, P_{j} \, P_{j+1} \, P_{j+2} \, \right) \\ & \quad \left( \, P_{j-3} \, P_{j-2} \, P_{j-1} \, \right) \\
    & = P_{j-3} \,  P_{j-2} \, P_{j-1} \, \sigma_{j}^+ \, P_{j+1} \, P_{j+2}  = P_{j-3} \ P_{j-2} \ \tilde{\sigma}_j^+ \ P_{j+2} 
\end{align*}
\newline
$(iv)$
\begin{align*}
    & \tilde{\sigma}_j^+ \  \left( \, \tilde{\sigma}_{j-1}^- \ \tilde{\sigma}_{j-1}^+ \, \right) \left( \, \tilde{\sigma}_{j}^- \  \tilde{\sigma}_{j}^+ \, \right)  \left( \, \tilde{\sigma}_{j-2}^- \  \tilde{\sigma}_{j-2}^+ \, \right)\  \\ &= \left(\, P_{j-1} \, \sigma_j^+ \, P_{j+1} \, \right) \ \left(\, P_{j-2} \, P_{j-1} \, P_{j} \, \right)  \ \left( \, P_{j-1} \, P_{j} \, P_{j+1} \, \right) \\ & \quad \left( \, P_{j-3} \, P_{j-2} \, P_{j-1} \, \right) \\
    & = P_{j-3} \,  P_{j-2} \, P_{j-1} \, \sigma_{j}^+ \, P_{j+1} = P_{j-3} \ P_{j-2} \ \tilde{\sigma}_j^+ 
\end{align*}
Because of the property that $\left( \, \tilde{\sigma}_{k}^- \ \tilde{\sigma}_{k}^+ \, \right)^p =  \tilde{\sigma}_{k}^- \ \tilde{\sigma}_{k}^+ $  for integer $p$, all the terms of $m=1$ in $(n-2)$-th order reappear at $m=1$ in $n$-th order. In particular, other additional terms here are $\tilde{\sigma}_{j}^+ \ , \ \, \tilde{\sigma}_{j}^+ \ P_{j+2} \ , \ P_{j-2} \ \tilde{\sigma}_{j}^+$ and $P_{j-2} \ \tilde{\sigma}_{j}^+ \ P_{j+2}$. Additionally in $7$-th order, for $m=3 \ , \ q = 2$, the following term is of interest. \\
$(i)$
\begin{align*}
    & \quad \tilde{\sigma}_{j-2}^+ \  \tilde{\sigma}_{j}^+ \,  \tilde{\sigma}_{j+2}^+ \ \left( \, \tilde{\sigma}_{j-1}^- \ \tilde{\sigma}_{j-1}^+ \, \right)  \left( \, \tilde{\sigma}_{j+1}^- \  \tilde{\sigma}_{j+1}^+ \, \right) \\ &= \left(\, P_{j-3} \, \sigma_{j-2}^+ \, P_{j-1} \, \right) \left(\, P_{j-1} \, \sigma_{j}^+ \, P_{j+1} \, \right) \left( \, P_{j+1} \, \sigma_{j+2}^+ \, P_{j+3} \, \right) \\  & \qquad\left(\, P_{j-2} \, P_{j-1} \, P_{j} \, \right) \left( \, P_{j} \, P_{j+1} \, P_{j+2} \, \right) \\ & = P_{j-3} \,  \sigma_{j-2}^+ \, P_{j-1} \, \sigma_{j}^+ \, P_{j+1} \, \sigma_{j+2}^+ \, P_{j+3} = \ \tilde{\sigma}_{j-2}^+ \ \tilde{\sigma}_{j}^+ \ \tilde{\sigma}_{j+2}^+
\end{align*}
This exercise justifies that terms in $H_F$ that lead to Fock states such as $\ket{\downarrow \cdots \downarrow \uparrow \downarrow \uparrow \downarrow \uparrow}$ from $\ket{\mathrm{vac}}$ can only appear at $7$-th order and higher in FPT. 

\begin{figure}  \label{figSM3}
     \includegraphics[width=\linewidth]{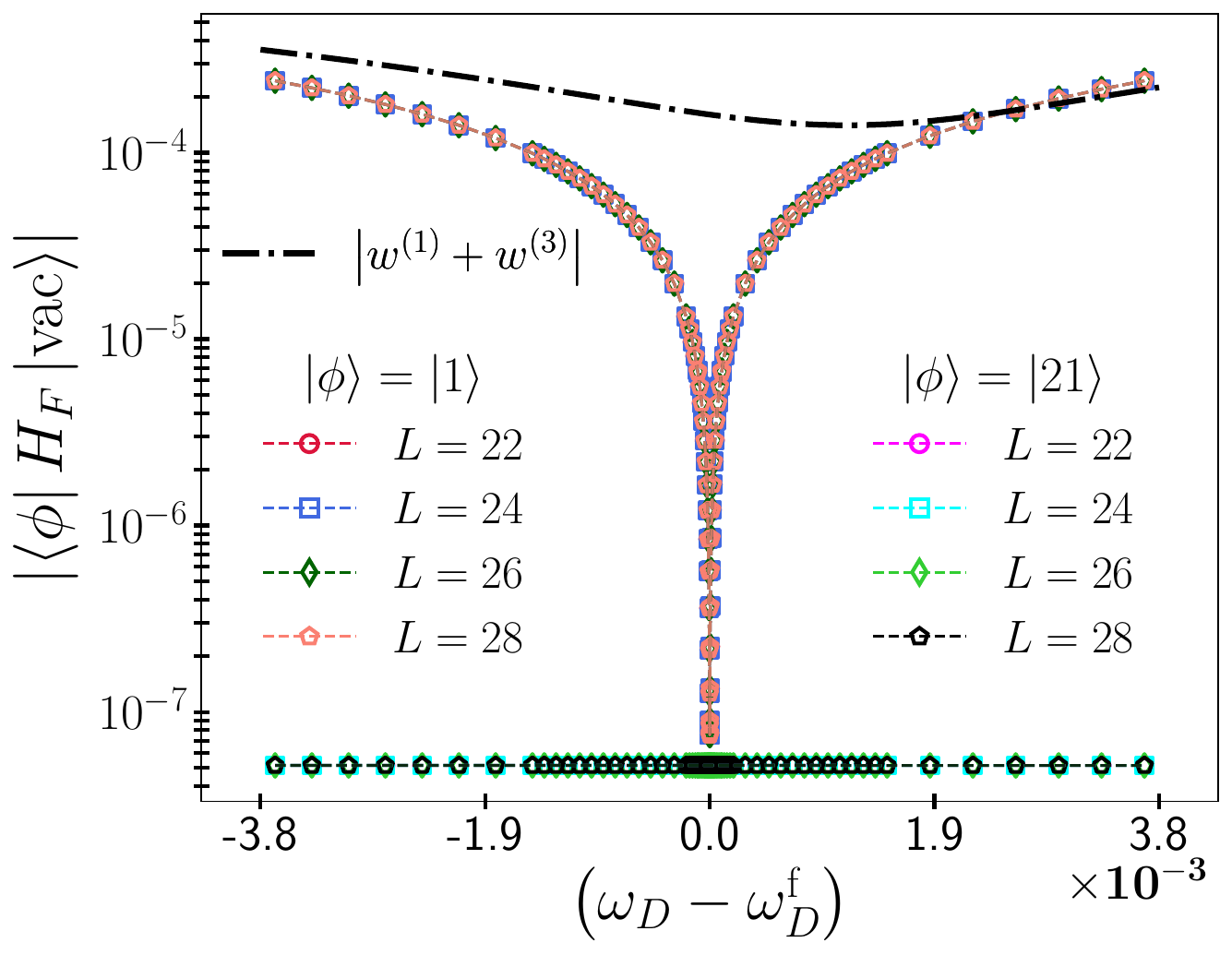}
     \caption{Magnitudes of matrix elements $\bra{\phi}H_F\ket{vac}$ near the drive frequency $\omega_D^{\mathrm{f}}\approx 15.0994496$ at $\lambda=30$ as a function of $\omega_D$ for the system sizes $L \in [22,24,26,28]$. Here the states $\ket{1}$ and $\ket{21}$ are the Fock states $\ket{\downarrow\cdots \downarrow \uparrow}$, and $\ket{\downarrow\cdots \downarrow \uparrow \downarrow \uparrow \downarrow \uparrow}$, respectively. The behavior of the magnitude of the matrix elements with $L$ shows a convergence to $L$-independent values. The magnitude of $w_{\mathrm{eff}}= w^{(1)}+w^{(3)}$ (see main text) obtained from the $3$rd-order FPT calculation is shown by the black dashed curve. 
     }
 \end{figure}

Now consider terms that generate Fock states with $m$ up spins starting from the vacuum at the lowest possible order $n$. This is achieved by placing lone operators at $m$ next-nearest-neighbor sites and then inserting $(m-1)$ pairs between these lone sites to form the smallest close-packed connected unit allowed by the strong Rydberg constraints. This fixes $q=m-1$ in Eq.~\ref{eqCons} and leads to 
\begin{equation}
    n_l = m+2(m-1)=3m-2
    \label{eqlowestorder}
\end{equation}
where $n_l$ is the lowest order in FPT at which such Fock states can be generated, with the restriction that $m$ is odd. This immediately shows that Fock states with up-spins such as $\ket{\downarrow \cdots \downarrow \uparrow \downarrow \uparrow \downarrow \uparrow \downarrow \uparrow \downarrow \uparrow}$ can be generated only from $ 13$-th-order and higher terms in FPT. The unusually high order of the terms that do not annihilate $\ket{\mathrm{vac}}$ within FPT, coupled with the significant renormalization of the magnitude of $w_{\mathrm{eff}}$ (see Fig.~\ref{figSM3} to be discussed in Sec.~\ref{sec_SM_F}) near $\omega_D^{\mathrm{f}}$, explains the exceptional stability of $\ket{\mathrm{vac}}$ near $\omega_D=\omega_D^{\mathrm{f}}$ for large values of $\lambda$. 

 \begin{figure} \label{figEM2}
    \includegraphics[width=\linewidth]{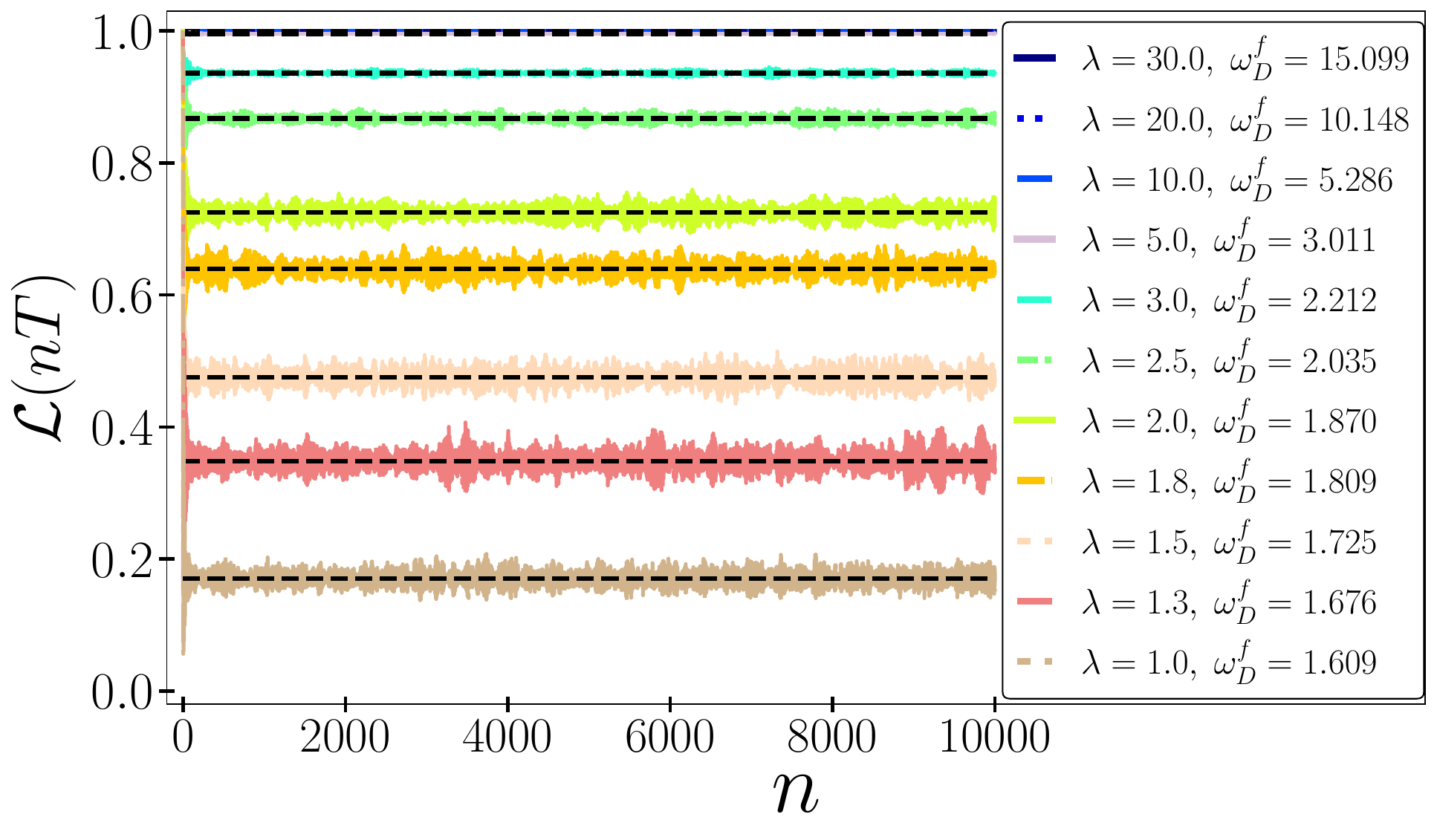}
    \caption{Loschmidt echo $\mathcal{L}(nT):= |\bra{\mathrm{vac}}U(T,0)^n\ket{\mathrm{vac}}|^2$ of the Rydberg vacuum state up to $n=10^4$ stroboscopic steps for some $(\lambda, \ \omega_D^{\text{f}})$, for system size $L=28$. The corresponding value of $W_0^2$ for the dressed vacuum scar at each $(\lambda, \ \omega_D^{\text{f}})$ is shown by horizontal black dashed lines. 
    } 
\end{figure}

\begin{figure}  \label{figSM4}
 \includegraphics[width=\linewidth]{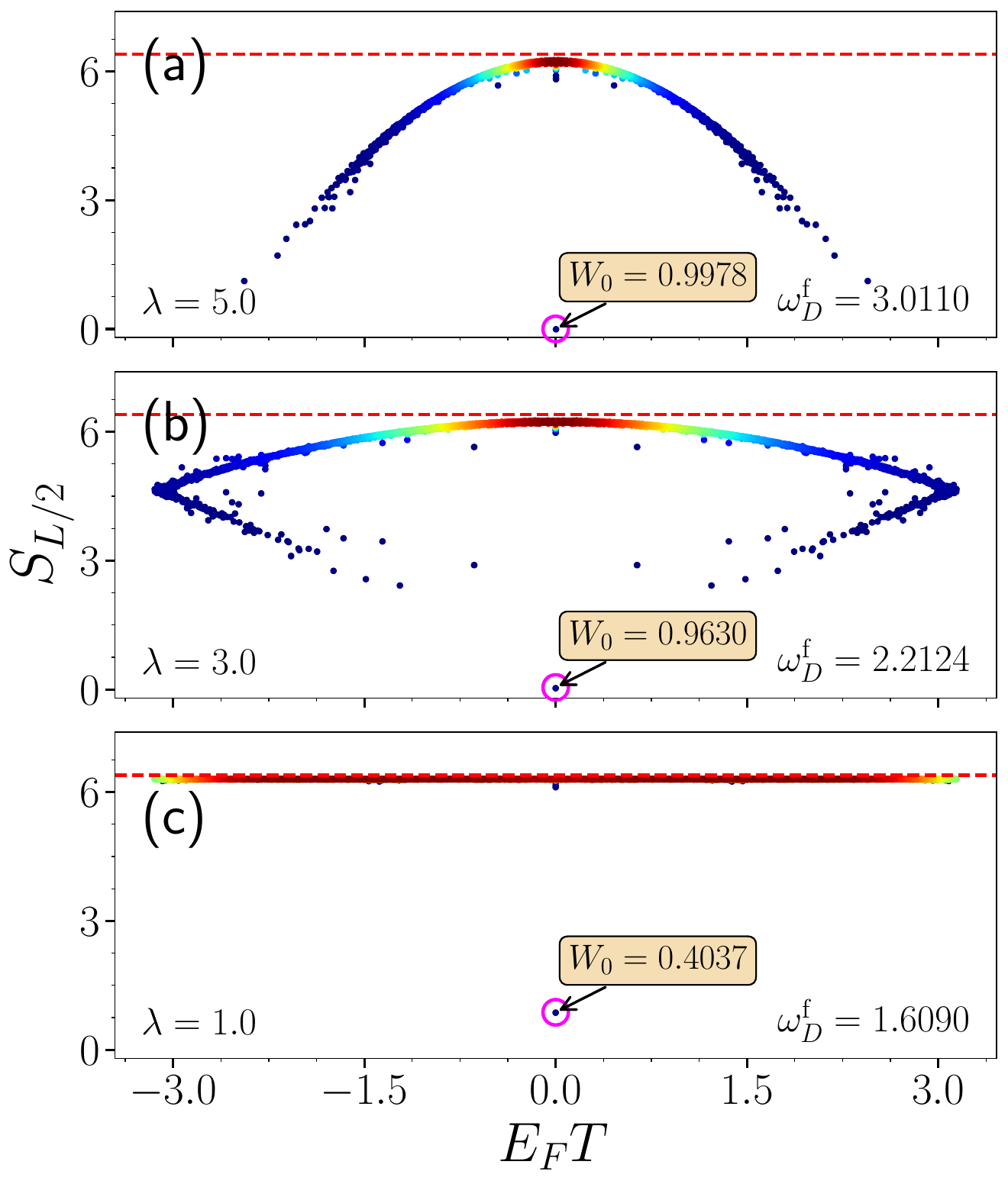}
 \caption{Plot of the half-chain entanglement entropy $S_{L/2}$ for eigenstates of $H_F$, where $E_F T \in [-\pi,\pi)$, at (a) $\lambda=5, \, \omega_D^{\mathrm{f}} =3.011$ (b) $\lambda=3, \, \omega_D^{\mathrm{f}} =2.2124$ and (c) $\lambda=1, \, \omega_D^{\mathrm{f}} =1.609$ for $L=28$ ($K=0,\mathcal{P}=+1$ sector).  The dressed vacuum scar is circled in magenta, and all panels show its overlap with the parent vacuum state, $W_0$. Note that $W_0 \gg W_0^{\mathrm{typ}} \simeq 0.02$ for all three panels. The Page value is approximately $6.394$ and is shown by a horizontal red dashed line in all the panels. The density of states is indicated by the same color map in all three panels, where warmer colors correspond to higher density of states. 
}
\end{figure}

\begin{figure} \label{figSM5}
 \includegraphics[width=\linewidth]{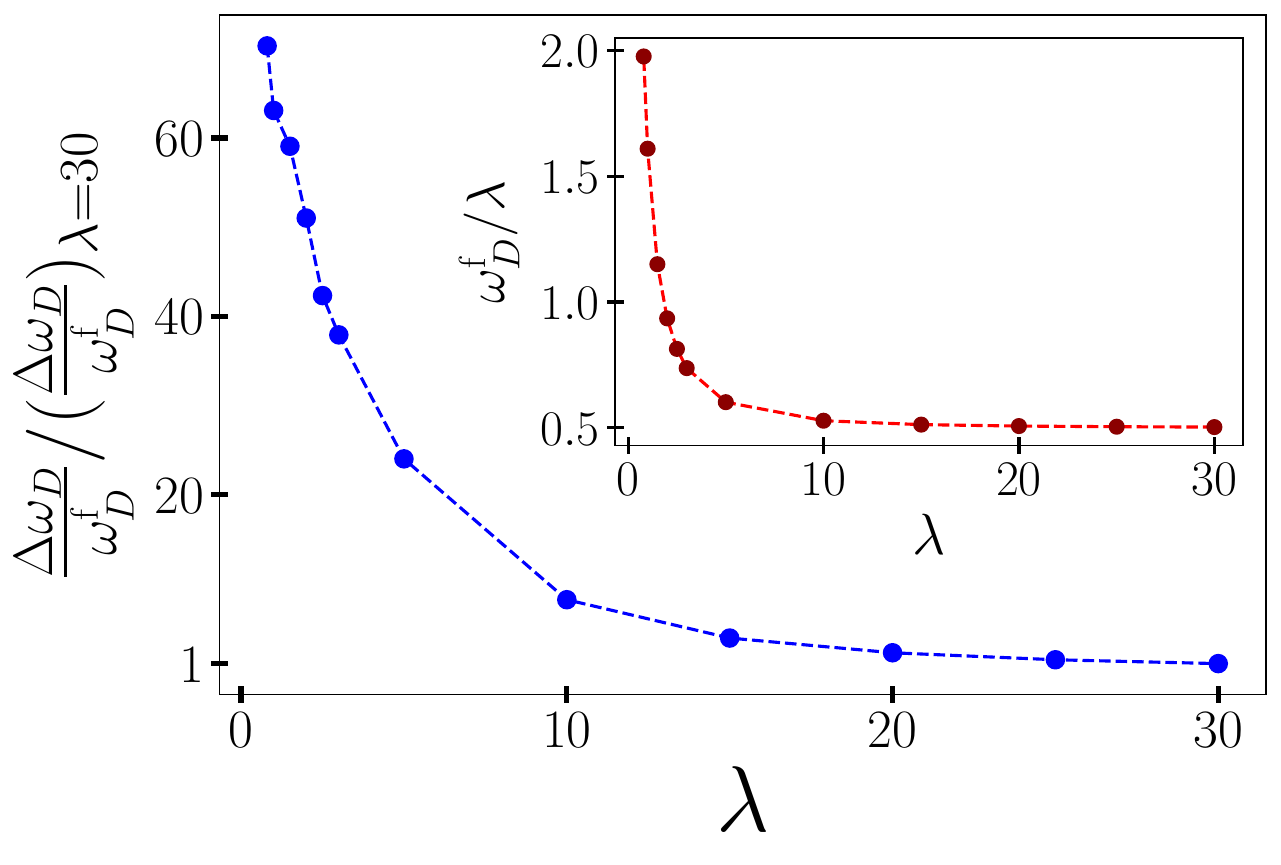}
 \caption{
 The ratio $\Delta(\omega_D)/\omega_D^{\mathrm{f}}$ (scaled by dividing with the value at $\lambda=30$) is shown as a function of the drive amplitude $\lambda$ for system size $L=26$. $\Delta(\omega_D)$ is the frequency window around $\omega_D^{\mathrm{f}}$ where $W_0(\ket{\rm vac})$ decreases by $10\%$ of its maximum value at $\omega_D^{\mathrm{f}}$. In the inset, $\omega_D^{\mathrm{f}}/\lambda$ is shown as a function of $\lambda$. 
 }
\end{figure}
\section{Additional features of the dressed vacuum scar}
\label{sec_SM_F}

We discuss additional numerical results showing nonperturbative features in the dressed vacuum scar at both large and small $\lambda$, beyond those mentioned in the main text. Since $H_F$ anticommutes with the chiral operator $\mathcal{C}$, any term appearing at arbitrary order in FPT (\ref{sec_SM_B}) must connect $|{\rm vac}\rangle$, a $\mathcal{C}=+1$ eigenstate, to other $\mathcal{C}=-1$ eigenstates. The latter states must be Fock states (or a superposition of them) that consist of an odd number of up-spins. Due to the hierarchy of the FPT terms, the dominant connection of $\ket{\mathrm{vac}}$ through $H_F$ is to the state $\ket{\downarrow \cdots \downarrow \uparrow}$ and all its translated partners in the perturbative regime. 

Fig.\ \ref{figSM3} displays the magnitude of the matrix element $\bra{\phi}H_F\ket{\mathrm{vac}}$ as a function of drive frequency near the first freezing point at $\lambda=30$ for the Fock states $\ket{\phi}=\ket{\downarrow \cdots \downarrow \uparrow}$ and $\ket{\phi}=\ket{\downarrow \cdots \downarrow \uparrow \downarrow \uparrow  \downarrow \uparrow}$, which essentially probes the renormalized one- and three-spin flip coefficients that are present in $H_F$. Although the two matrix elements are separated by several orders of magnitude for a generic drive frequency in the perturbative regime $\lambda \gg 1$, the magnitude of the renormalized single-spin flip term $(\approx 7.53 \times 10^{-8})$ becomes comparable to the magnitude of the three-spin flip term ($\approx 5.16 \times 10^{-8}$) at $\omega_D^{\mathrm{f}}$. Fig.~\ref{figSM3} also shows the magnitude of the renormalized single-spin flip coefficient $w_{\mathrm{eff}}$, analytically computed up to 3rd order in FPT (see the main text and Sec. \ref{sec_SM_B} for the expression). It turns out that $|w_{\mathrm{eff}}|$ does not have a minimum at the freezing frequency $\omega_D^{\mathrm{f}}$ and is not symmetric about it. More importantly, $|w_{\mathrm{eff}}|$ is approximately $\mathcal{O}(10^{-4})$ near $\omega_D^{\mathrm{f}}$ and, for the range of frequency specified in the figure, it fluctuates between $10^{-3}- 10^{-4}$. This shows a strong renormalization of the single-flip coefficient from higher-order FPT terms even at $\lambda=30$, and results in the single-spin-flip coupling being as low as $10^{-8}$, as observed numerically at $\omega_D^{\mathrm{f}}$.

Next, we monitor the return probability (Loschmidt echo) $\mathcal{L}(nT):= |\bra{\mathrm{vac}}U(T,0)^n\ket{\mathrm{vac}}|^2$ (Fig. \ref{figEM2}) as a function of $n$ for specific values of $\lambda$. For these plots, we tune the drive frequency to $\omega_D^{\mathrm{f}}(\lambda)$ so that $W_0(|\mathrm{vac} \rangle)$ is maximized at each $\lambda$. From Fig.\ \ref{figEM2}, we find that the Loschmidt echo hovers around $1$ for $\lambda \geq 5$ as a function of $n$ for $L=28$. This is expected since $W_0(|\mathrm{vac} \rangle)$ is very close to $1$ in this regime. In contrast, the runs for $\lambda \in [1.0, 3.0]$ (Fig.~\ref{figEM2}) indicates that $\mathcal{L}(nT)$ shows an initial rapid decay from $1$. It then fluctuates slightly around $W_0^2$ (shown as dotted black lines) in each case. Although $W_0$ decreases as $\lambda$ decreases, it remains finite in all cases. Using spectral decomposition
\begin{align}
|\psi \rangle &= \sqrt{W_0(\psi)} |Z_{\psi} \rangle+ \sum_{E_i>0}  \left[ a(E_i) \ket{E_i}  + a(-E_i)  \ket{-E_i}  \right] \ ,
\end{align}
we can calculate the long-time average of the Loschmidt echo for a parent state $\ket{\psi}$, which yields
\begin{equation} \label{eq_return_prob}
\begin{aligned}
\langle \mathcal{L} \rangle &= \lim_{n_0 \to \infty} \frac{1}{n_0} \sum_{n=0}^{n_0} \ \mathcal{L}(nT) \\
    &= W_0 (\psi)^2  +  \sum_{E_i>0} \Big( |a(E_i)|^4 + |a(-E_i)|^4 \Big).
\end{aligned}
\end{equation}
Thus, $\langle \mathcal{L} \rangle \geq W_0^2$ and is well approximated by $W_0^2$ when $|\mathrm{vac} \rangle$ has little overlap with any $E \neq 0$ modes of $H_F$. Thus $\mathcal{L}(nT)$ fluctuates around $W_0^2$ for all the cases displayed in Fig.~\ref{figEM2}, including the non-perturbative regime where $\lambda \sim O(1)$. This feature is consistent with the stability of the dressed vacuum scar in the non-perturbative regime discussed in the main text. 

Fig.\ \ref{figSM4} shows plots of the half-chain entanglement entropy, $S_{L/2}$, for Floquet eigenstates. $S_{L/2}$ is calculated using the formula  $S_{L/2}=-\mathrm{tr(\rho \ln \rho)}$ with $\rho$ being the reduced density matrix for $\omega_D^{\mathrm{f}}$ at three different values of $\lambda$. Here we have obtained the ED data for a system of size $L=28$ using the $K=0, \mathcal{P}=+1$ sector. A typical mid-spectrum eigenstate is expected to have $S_{L/2}$ close to the Page value,
which can be obtained using the expression $\ln{\mathcal{D}_{L/2}} - 1/2$, with $\mathcal{D}_{L/2}$ being the Hilbert space dimension of $L/2$ spins ($L$ is even) under OBC. 

From Fig.\ \ref{figSM4}, we find that the dressed vacuum scar has an anomalously low entanglement entropy for all three values of $\lambda$. The behavior of $S_{L/2}$ for the $E_F \neq 0$ Floquet eigenstates as a function of $\lambda$ clearly shows the evolution from a regime where $H_F$ can be approximated to be a local Hamiltonian (Fig. \ref{figSM4} (a) at $\lambda=5.0$) to one where $H_F$ is completely nonlocal (Fig. \ref{figSM4} (c) at $\lambda=1.0$), consistent with the behavior of $\langle S^z \rangle$ shown in the main text. While $W_0$ of the dressed vacuum scar decreases significantly as $\lambda$ changes from $5.0$ to $1.0$ (Fig. \ref{figSM4} (a), (b), (c)), the dressed vacuum scar remains an outlier both with respect to $S_{L/2}$ as well as the local operator $\langle S^z \rangle$ (main text).  

In Fig.~\ref{figSM5}, we show the behavior of the width $\Delta \omega_D$ around the drive frequency $\omega_D^{f}$ as a function of $\lambda$ for $L=26$. To find $\Delta \omega_D$, we calculate $W_0(\ket{\rm vac})$ and identify the window of drive frequencies where it reaches $0.9$ times its maximum value at $\omega_D^{\mathrm{f}}$. A plot of the dimensionless quantity $\Delta \omega_D / \omega_D^{\mathrm{f}}$ as a function of $\lambda$ shows that this width increases as $\lambda$ is lowered to $O(1)$ values. The enhancement in $\Delta \omega_D$ around $\omega_D^{f}$ is sizable as $\lambda$ is decreased from $\lambda=30$ to $\lambda=0.8$ as is evident from Fig.~\ref{figSM5}. The inset of Fig.~\ref{figSM5} shows the dependence of $\omega_D^{f}/\lambda$ on $\lambda$ for the same system size. As the inset shows, $\omega_D^{f}/\lambda$ approaches $1/2$ for large $\lambda$ (see main text), but deviates substantially from $1/2$ as $\lambda$ approaches $O(1)$. Both Fig.~\ref{figSM4} and Fig.~\ref{figSM5} illustrate how features of the dressed vacuum scar, including its domain of stability in drive frequency $\omega_D$, evolve as $\lambda$ is varied from the perturbative regime ($\lambda \gg 1$) to a nonperturbative one ($\lambda \sim O(1)$).   

